\documentclass[graybox]{svmult}

\usepackage{amssymb}
\usepackage{amsmath}
\usepackage{mathptmx}       
\usepackage{helvet}         
\usepackage{courier}        
\usepackage{type1cm}        
\usepackage{tabularx}
\usepackage{color}
\usepackage{subeqnarray}
\usepackage{makeidx}         
\usepackage{graphicx}        
\usepackage{multicol}        
\usepackage[bottom]{footmisc}

\newcommand{\be}{\begin{equation}}
\newcommand{\ee}{\end{equation}}
\newcommand{\bae}{\begin{eqnarray}}
\newcommand{\eae}{\end{eqnarray}}
\newcommand{\bse}{\begin{subeqnarray}}
\newcommand{\ese}{\end{subeqnarray}}

\newcommand{\tabref}[1]{Table~\ref{tab:#1}}
\newcommand{\figref}[1]{Figure~\ref{fig:#1}}

\newcommand{\Exp}[1]{\ensuremath{e^{#1}}}
\newcommand{\Abs}[1]{\ensuremath{\left|#1\right|}}
\renewcommand{\D}[2]{\ensuremath{\frac{\partial {#1}}{\partial {#2}}}}
\newcommand{\Du}[2]{\ensuremath{\frac{\textup{d} {#1}}{\textup{d} {#2}}}}

\newcommand{\bvec}[1]{\ensuremath{\mathbf{#1}}}

\newcommand{\PIA}{Pleistocene Ice Ages~}

\newcommand{\Ecce}{eccentricity~}
\newcommand{\Obli}{obliquity~}
\newcommand{\Prec}{precession~}

\makeindex             

\begin{document}

\title*{Numerical Bifurcation Analysis of Marine Ice Sheet Models}
\author{T. E. Mulder, H. A. Dijkstra and F. W. Wubs}
\institute{T. E. Mulder and H. A. Dijkstra \at Institute for Marine and Atmospheric research Utrecht, 
Department of Physics, Utrecht University, the Netherlands, \email{h.a.dijkstra@uu.nl}
\and F. W. Wubs  \at Johann Bernoulli Institute for Mathematics and Computer Science, 
Groningen University, Groningen, The Netherlands
 \email{f.w.wubs@rug.nl}}
%
%
\maketitle

\abstract{The climate variability associated with the \PIA is one of the most  fascinating puzzles 
in the Earth Sciences still awaiting a satisfactory explanation. In particular, the explanation of 
the dominant 100 kyr period of the glacial cycles over the last million years is a long-standing  
problem.  Based on bifurcation analyses of low-order models, many  theories have been suggested 
to explain these cycles and  their frequency. The new aspect in this contribution  is that,  for the first  time, 
numerical bifurcation analysis is  applied to a two-dimensional marine ice sheet model with a dynamic
grounding line. In this model, we find Hopf bifurcations with an oscillation period of about 100 kyr 
which may be relevant to glacial cycles. } 
\vspace{1cm} 

\section{Introduction}

Very detailed information on past temperatures on Earth has been obtained from marine benthic 
records, in particular oxygen isotope ratios.   Water in ice cores contains two  isotopes of oxygen, 
$^{18}$O and $^{16}$O.   The normalized isotope ratio   $\delta^{18}$O is calculated as a 
deviation from a  reference sample as 
 \be                                                                                                                                                                                                                                                                                                                                                                                                                                                                                                                                                                                                                                                                                                                                                                                                                                                                                                                                                                                                                                                                                                                                                                                                                                                                                    
 \delta^{18}O = \frac{(\frac{^{18}O}{^{16}O})_{sample} -  
                (\frac{^{18}O}{^{16}O})_{reference}}
               {(\frac{^{18}O}{^{16}O})_{reference}}, 
 \ee
 where the reference sample is `standard mean' ocean water.   The isotope  $^{16}O$ 
 is lighter  than $^{18}O$ so that water containing  $^{16}O$ is preferentially 
 evaporated   and a temperature-dependent fractionation occurs.  Changes in 
 $\delta^{18}O$ reflect the combined effect of changes in global ice volume and 
 temperature at the time of deposition of the sampled material.   During very cold conditions, 
 global  ice volume is relatively large and hence sea level is low, which enriches water in the 
 ocean with $^{18}O$. Also because of the colder temperatures, more  $^{18}O$ remains   
 in the ocean and less  $^{18}O$ becomes  locked in the ice.  Hence, the  ratio 
 $\delta^{18}O$ in ice cores  will decrease (becomes more negative) under colder 
 conditions.  
 
\begin{figure}
\begin{minipage}[hbtp]{\linewidth}
\centering\includegraphics[width=\linewidth]{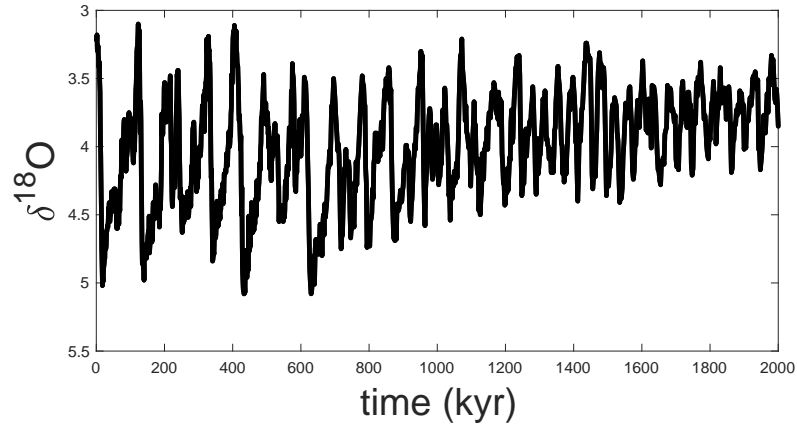} 
\end{minipage} 
\caption{\em (a) The LR04 benthic $\delta^{18}O$   stack over the Pleistocene, 
constructed by the graphic  correlation of 57 globally distributed benthic 
$\delta^{18}O$ records (data from \cite{Lisiechi2005}). The x-axis indicates 
time BC in kyr (so the time is increasing from right to left).  
} 
\label{f:Fig4Lisiecki}
\end{figure}

 In marine sediment cores the opposite behavior of the isotope ratio 
 is found  and $\delta^{18}O$ increases (becomes more positive) when the temperature decreases 
(during colder  conditions, the concentration of the heavier isotope will increase). 
In Fig.~\ref{f:Fig4Lisiecki}, a 
time series is shown  of a composite $\delta^{18}$O ocean sediment (benthic) record over 
the last  2  Myr \cite{Lisiechi2005}.    A  cooling  trend is found on which variability in ice cover  is 
superposed.  Analysis reveals that this variability is first dominated by a 41 kyr  period  
and after the so-called Mid Pleistocene Transition (MPT) at  about 700 kyr, it is   dominated by  
a 100 kyr period. 

\begin{figure}
\begin{minipage}[hbtp]{\linewidth}
\centering\includegraphics[width=0.75\linewidth]{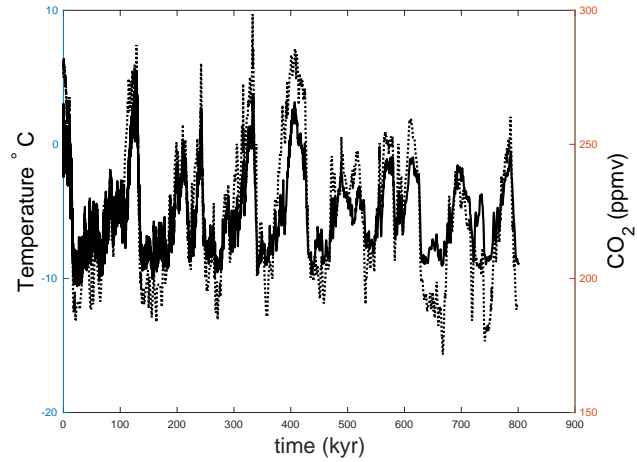} 
\end{minipage} 
\caption{\em Reconstructed temperature (drawn) and atmospheric CO$_2$ (dotted) 
concentration  from ice cores on Antarctica. 
}  
\label{f:Fig2Jouzel}
\end{figure}

The European  Project for Ice Coring in  Antarctica (EPICA) has provided two deep ice 
cores in East  Antarctica  from which climate conditions can be reconstructed back to 
800 kyr BP \cite{Jouzel2007}.  From  the reconstructed temperature anomaly
time series (Fig.~\ref{f:Fig2Jouzel}),  one  observes the  asymmetry  
between the slow glaciation and the rapid deglaciations. 
Glacial-interglacial transitions have affected all components of the 
climate system and induced relatively large amplitude changes of many variables
in these components.  One important  player in the climate system responsible for 
the globalization of these transitions 
is believed to be the atmospheric CO$_2$ concentration.   A composite CO$_2$ record  
is also  shown in Fig.~\ref{f:Fig2Jouzel}, created 
from a combination of records from the Dome C  and Vostok ice cores. 
It is observed that the atmosphere CO$_2$ concentration 
varies from about 180 ppm to 280 ppm during
a glacial-interglacial transition and that an optimal correlation  with  the  $\delta^{18}$O 
time series occurs near lag zero. 

These results lead to many intriguing questions: Why did  glacial-interglacial 
cycles appear in the Pleistocene? 
Which processes in the climate system caused  the glacial-interglacial 
changes in global mean temperature   and ice sheet extent? 
What caused the transition (the MPT) from the 41 kyr world to 
the 100 kyr world about 700 kyr ago? 

Approaches to answers  on the \PIA problem have a very  interesting history 
which is nicely described in \cite{Imbrie1986}.  
A connection with 
the orbital characteristics of the Earth-Sun system was already made in the 19$^{th}$ 
century,  but  in the 1930s it was suggested \cite{Milankovich1930}  
that glaciations occur when the insolation intensity is weak at high northern  latitudes 
during summer.  When  the 65$^\circ$N  insolation is small,  ice can persist throughout the 
year leading to the growth of  ice  sheets.  Favorable conditions for this to happen are 
when the spin axis is less tilted and the aphelion (the point in the orbit, where 
the Earth is farthest from the Sun) coincides with summer in the Northern 
Hemisphere. 

\begin{figure}
\begin{minipage}[hb]{0.45\linewidth}
\centering\includegraphics[width=\linewidth]{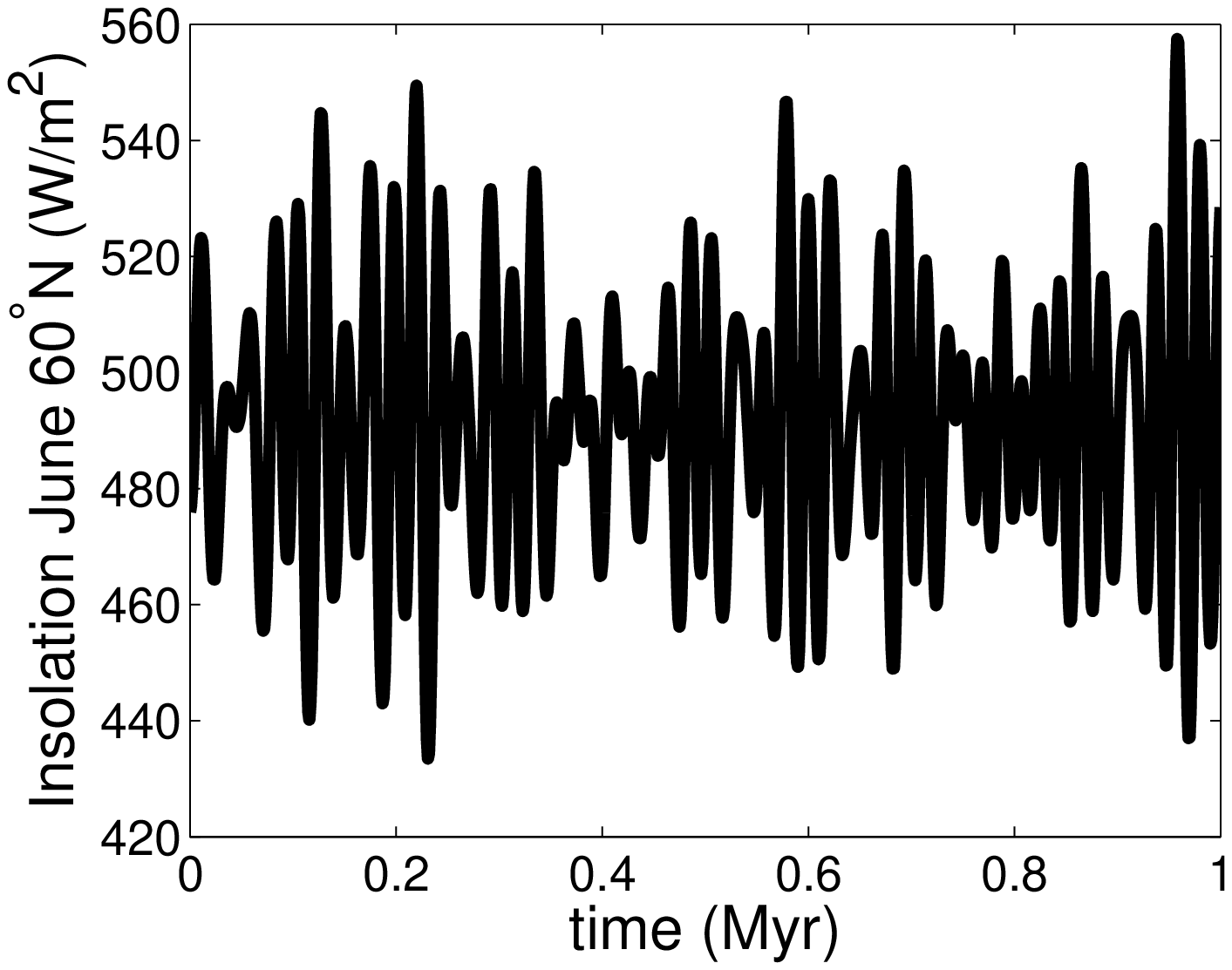} (a) 
\end{minipage} 
\begin{minipage}[hb]{0.45\linewidth}
\centering\includegraphics[width=\linewidth]{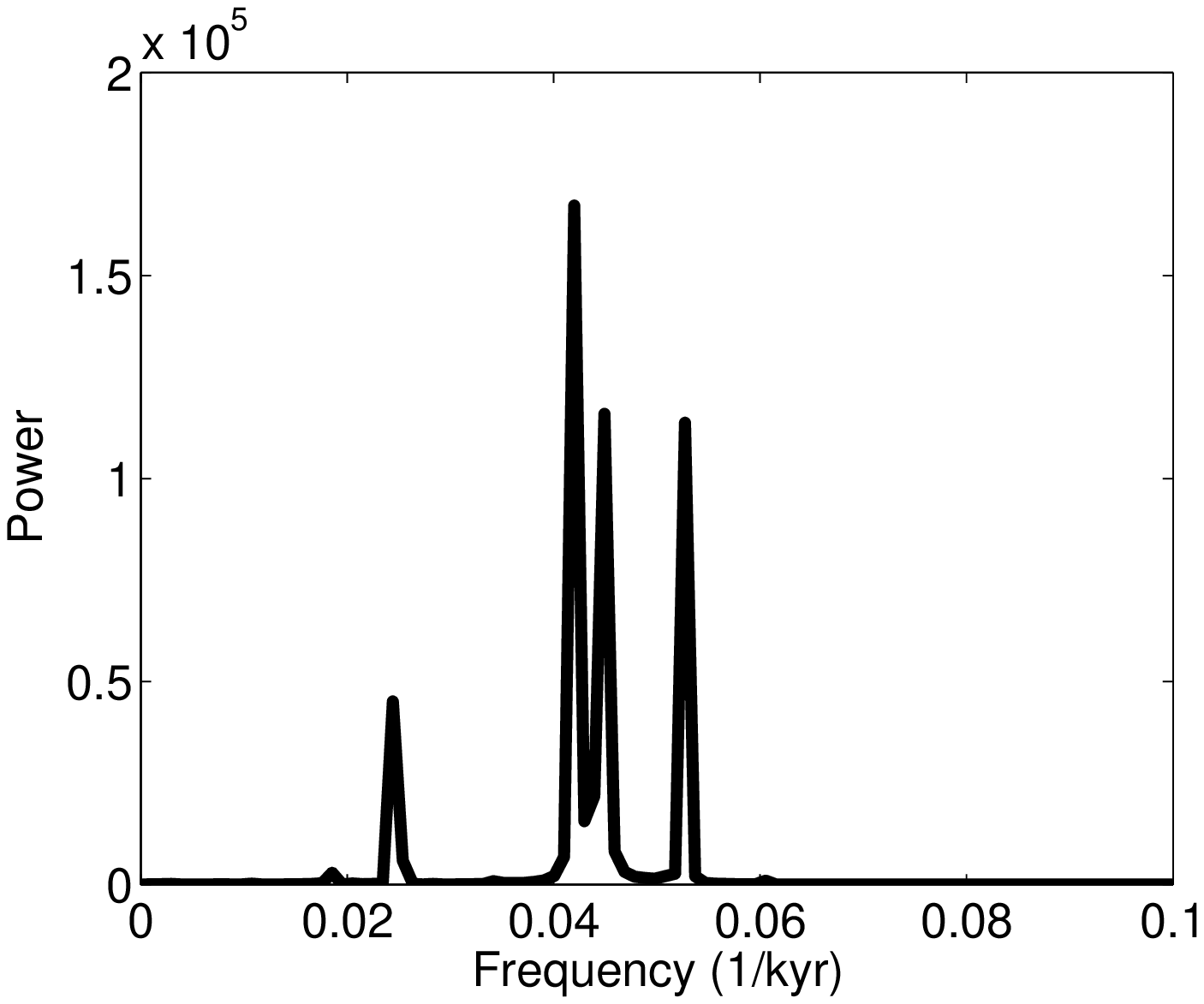} (b) 
\end{minipage} \\
\begin{minipage}[hb]{0.45\linewidth}
\centering\includegraphics[width=\linewidth]{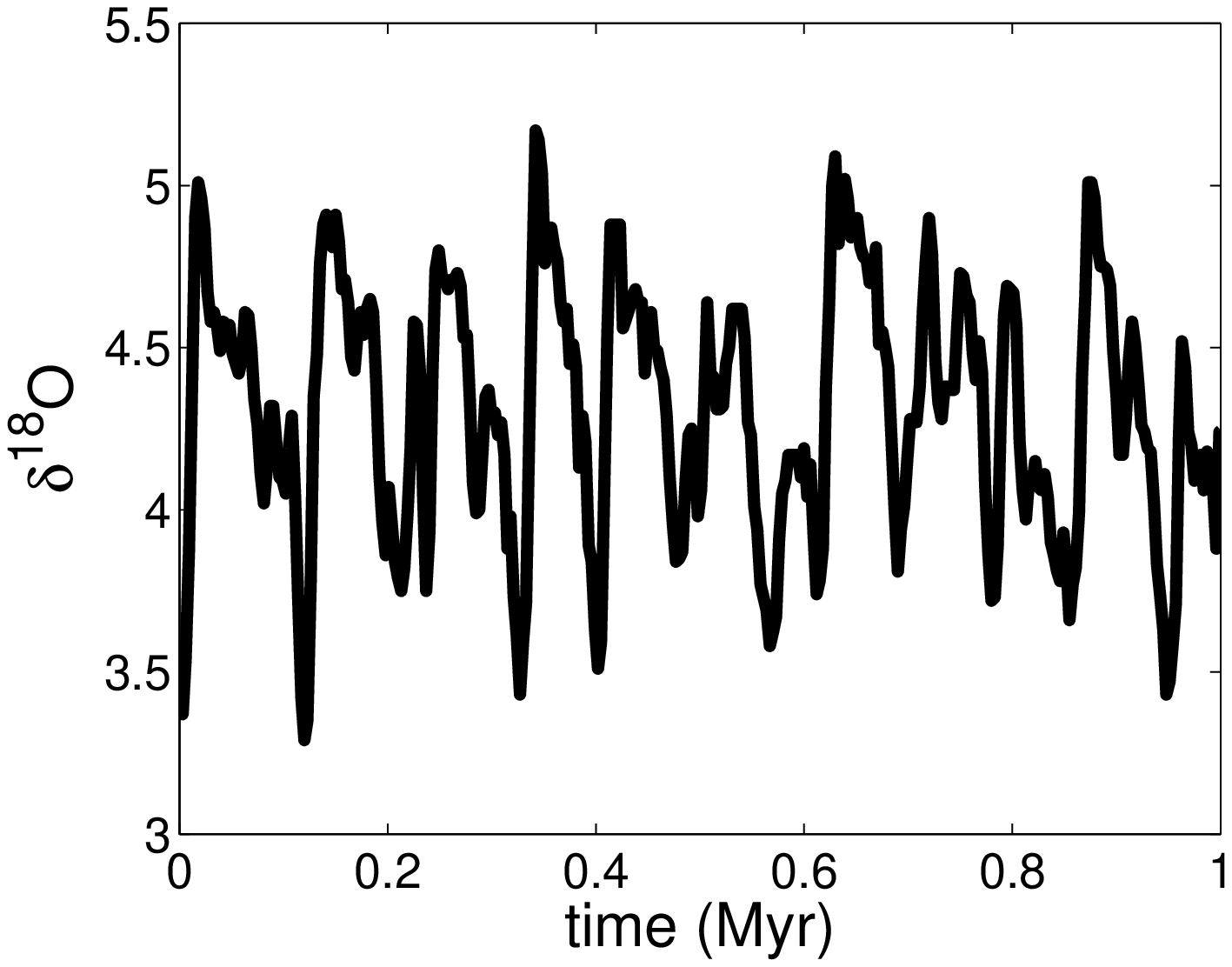} (c) 
\end{minipage} 
\begin{minipage}[hb]{0.45\linewidth}
\centering\includegraphics[width=\linewidth]{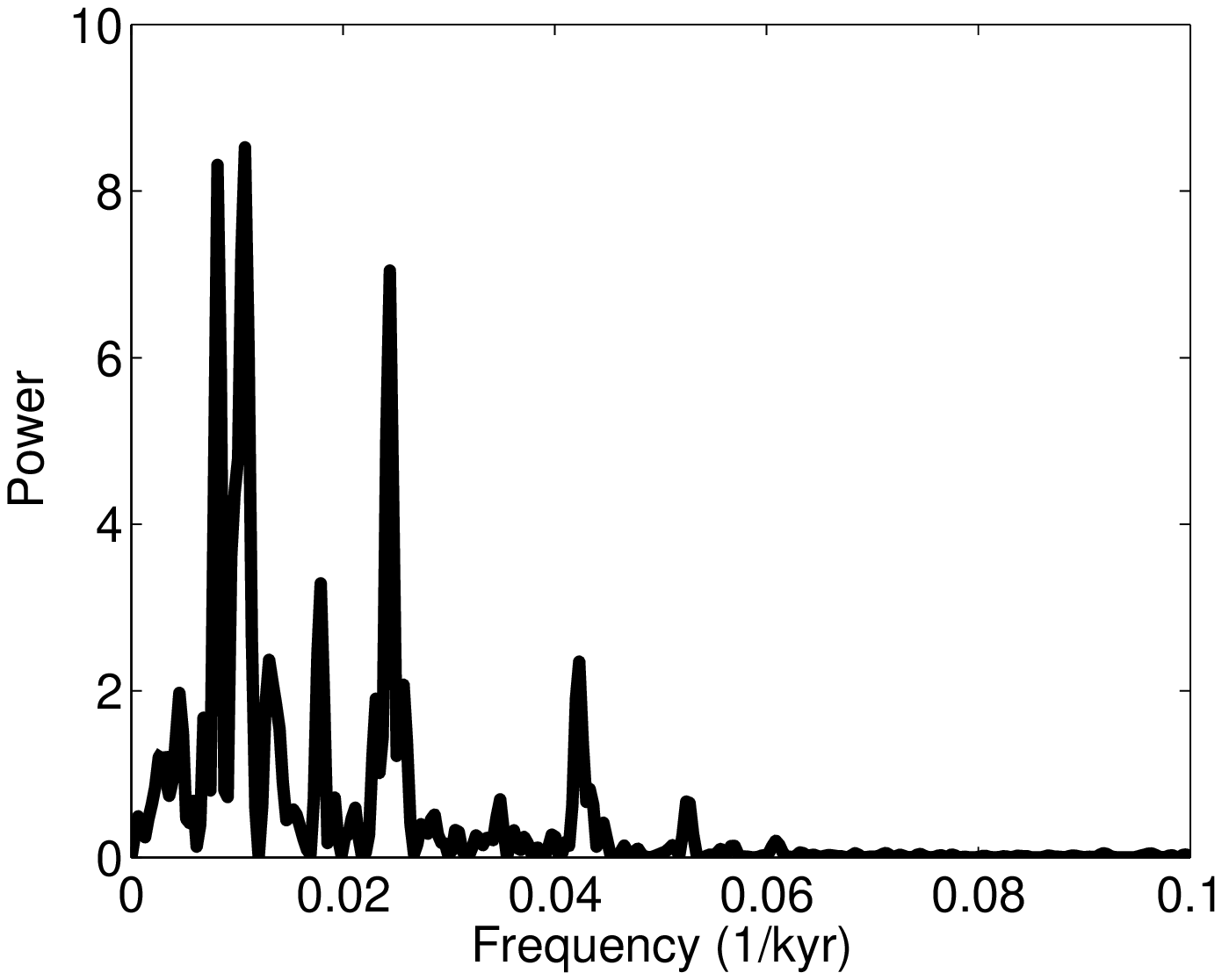} (d) 
\end{minipage} 
\caption{\em  (a) June insolation at 60$^\circ$ N. (b) Spectrum of the time series 
in (a). (c) Time series of $\delta^{18}O$ at ODP677 (83$^\circ$W, 1$^\circ$N)
over the last 1 Myr. (d)  Spectrum of the time series in (c). Note that $1/23 = 0.043$, 
$1/19 = 0.053, 1/41 = 0.024$ and $1/100 = 0.01$. 
} 
\label{f:spectra}
\end{figure}

The variations in insolation are caused by the changes in  orbital characteristics of 
the Earth,  and  there are three types of motion relevant for the amount  of radiation 
received at a  particular point on Earth. First, the spin axis of the   Earth undergoes  
\index{orbital parameters!precession}  precession.  One full cycle of precession has 
a period of 27 kyr,   but coupled to the movement of the long end of the ellipse 
around the Sun (in  105 kyr) the net effect is a fluctuation in solar radiation with a 
period of 23 kyr.   In addition, both the obliquity  and the eccentricity  of the Earth's 
orbit undergo  periodic variations. The tilt angle changes in 41 kyr between 
22$^\circ$ and  24$^\circ$ leading to variations in seasonal contrast,  and the 
eccentricity varies from 0.0 (perfect circle) to about 0.05 with  periodicities of 
100 kyr and 450 kyr.  
A time series of the insolation at 60$^\circ$N (Fig.~\ref{f:spectra}a) clearly 
shows variations of about 100 Wm$^{-2}$ over the last 1 Myr. 
A comparison of the spectrum of this insolation curve (Fig.~\ref{f:spectra}b)  
and  the  spectrum of a $\delta^{18}O$ record from an ocean sediment core 
(Fig.~\ref{f:spectra}d)  over the last 1Myr (as shown in Fig.~\ref{f:spectra}c) shows 
that there are clear signatures of the 19 and 23 kyr precession  and of  the 41 
kyr obliquity variations of the Earth's orbit in the Ocean Drilling Program record. 
On the other hand,  at the 100 kyr time scale, there is hardly any forcing amplitude  
while the climate signal in the  $\delta^{18}O$ record has the largest amplitude. 
Comparing   $\delta^{18}$O  records with the insolation time series \cite{Raymo2008},
it is interesting that the amplitude of the ice-cover variations   can  be very large while the insolation 
variation is very small. Furthermore, the variations  in insolation provide no clue on the transition 
from the 41 kyr world to the 100 kyr world as they  have the same temporal characteristics  
through the transition. 

Although it is clear that the orbital insolation variations must play a 
role, a simple linear forcing-response relation apparently does not  apply. The 100 kyr variations in 
insolation due to eccentricity are very weak (and it is the only forcing with an annual mean
signal) and the 41 kyr and 23 kyr provide only low-frequency variations on the seasonal 
variations, not on the annual mean insolation. Hence, processes internal to the climate system must 
play a role in the amplification of the orbitally induced  insolation variations. 

\section{Basic theories of interglacial-glacial cycles}

There have been many suggestions  on the dominant mechanisms of ice-age 
variability.  The orbital  variations  in insolation over the globe are 
at the heart of all these theories. Traditionally, the June insolation 
at 65$^\circ$N (such as shown in Fig.~\ref{f:spectra}a) has been used 
as the most important part of this forcing  as this determines whether snow 
will be left  at the end of the Northern  Hemisphere Summer season. 
Below, we will refer to this orbital component of the insolation as the 
M-forcing.

In a first theory,  the behavior of the  climate system is  considered as a 
transient  deviation from a single steady equilibrium due to the M-forcing.
One can calculate  that a 1\% change in solar insolation  leads to about a  
1$^\circ$C temperature change.  The dominant  amplitude of \Ecce  variations 
is  at a period of about 400 kyr   while the next  strongest variations occur on 
about  a 100 kyr time scale.  The variations  in  
\Ecce  do modify the globally and annually-averaged  amount of insolation but the amplitudes  
are very small, in the order of 0.1\% of the solar  constant. The variations in \Ecce  can 
therefore only account for a climate signal of at  most 0.1 K.  A similar analysis gives that the 
total variations in \Prec and \Obli can only account directly for a signal  of at most  0.5$^\circ$C, 
about an order of magnitude too small. When considering many other processes  (e.g., ice sheets, 
bedrock) in the direct response to the M-forcing, the sensitivity does not increase enough to explain 
the climate signal \cite{Ghil1994b}. 

There have been several suggestions that the existence of multiple steady states gives, 
together with the M-forcing, rise to glacial cycles. For example, in the model 
by  \cite{Paillard1998} there are three  equilibrium states in the climate system: an 
interglacial state {\bf i},  a weak glacial state {\bf g}, and a strong glacial state 
{\bf G}. Transitions  from {\bf i} to {\bf g} occur when the summer insolation at 
65$^\circ$N  drops below a value $i_0$.  Furthermore, transitions  from  {\bf g} 
to {\bf G} occur when the ice volume $V$ increases above some critical level $V_c$. 
Finally, a transition from  {\bf G} to {\bf i}  occurs when the insolation increases above 
a level  $i_1$. These are the only transitions which are allowed in this model. 
By forcing the model with the M-forcing, there is a good overall agreement between 
model and observations (considering the simplicity of the model). By allowing for a 
slight linear  trend in the maximum ice volume and one in the insolation 
forcing, \cite{Paillard1998}  also finds the MPT  at around the correct time and  
the spectra  of his model and typical  $\delta^{18}O$ data correspond reasonably
well to each other. 

When multiple equilibria, a weak periodic forcing, and noise are present
there is also the possibility of stochastic resonance. In fact, the discovery of stochastic resonance 
actually occurred \cite{Benzi1982, Benzi1983, Nicolis1982} while 
trying to explain the 100 kyr dominant glacial cycles using an  energy 
balance atmosphere model. The central element of this 
theory is the  amplification of the weak  periodic \Ecce component of 
the M-forcing by noise in the presence of multiple equilibria. 

Coupled  atmosphere-cryosphere  and cryospheric-lithosphere processes can 
also give  rise to internal variability. The question is of course, whether these
processes can also give rise to sustained oscillations (through Hopf 
bifurcations). If so, the  time scales and amplitude ranges of these 
oscillations with `realistic' values of the parameters will be of interest
regarding the glacial cycles. Typical results can be found in \cite{Ghil1987}
where a sustained oscillation is indeed present  under steady forcing. The  
time scale of the oscillation is about 6-7 kyr and not 100 kyr which 
shows that the processes captured in this simple model are not able 
to generate this long time scale. 
In \cite{LeTreut1988}, the model in \cite{Ghil1981} is forced by an M-type 
forcing and the 100 kyr period  arises due to nonlinear 
resonances of the external  frequency of the forcing and the internal frequency 
of oscillation  \cite{Ghil1994b}. 

Many other idealized models have been proposed which involve other components 
of the climate system \cite{Saltzman2001}, for example, those involving  the 
atmospheric concentration of CO$_2$ and the global ocean state. The model proposed 
by \cite{Paillard2004} attributes to the Antarctic ice sheet extent a central role in linking 
climatic and CO$_2$ glacial-interglacial changes. The model proposed by \cite{Omta2013} 
investigates the role of marine calcifiers in glacial-interglacial cycles. For many of these 
models, bifurcation diagrams  have been computed showing that oscillations are 
associated with Hopf bifurcations \cite{Crucifix2012}. 

A relatively simple model, where an internal oscillation exists with 
a time scale of about 100 kyr is that of \cite{Gildor2000}. The model
is a box model  of the climate where a similar atmospheric-cryospheric
model as in the previous subsection is coupled to an ocean model 
with a sea-ice component.  In \cite{Gildor2001}, results of the model 
forced by constant annual mean 
insolation  (no seasonal or   Milankovitch forcing)  are presented 
to assess  the degree to which the internal processes (particularly 
sea-ice) may  control  glacial cycle variability. A typical result 
for (near) standard values of the parameters in  
\cite{Gildor2001} shows that oscillations with a time scale of 
about 100 kyr are found.  The proposed mechanism of the variability 
is referred to as the sea-ice switch, where  rapid sea-ice growth and 
decay can act as a  switch  for the precipitation-temperature feedback 
affecting the growth   and/or decay of ice sheets. 
The 100 kyr time scale is due to the growth and decay of ice sheets 
which is coupled by relatively rapid sea ice changes. 

When this model is forced with Milankovitch insolation changes,  nonlinear  
resonances may occur  between  the internal oscillation and the orbital  
forcing leading to time  series which  qualitatively resemble the observed 
records  \cite{Tziperman2006}, which demonstrates how phase locking to
Milankovitch forcing affects glacial cycles in this  idealized
model. 
These nonlinear resonances are likely  to be present  in  every model 
where a  strong nonlinear interaction is  represented, explaining for
example the good  agreement between  very conceptual models, 
where only a multiple state switch \cite{Paillard1998} is 
represented, and observations. In other  words, even if the mechanism 
of the glacial-interglacial variability  is incorrect, there may still be a 
good fit with the isotopic record.  Due to the synchronization,  a comparison 
between time  series of simple models and isotope records is not 
mechanistically  selective \cite{Crucifix2016}.  

Although many bifurcation studies have been done on low-order models, there 
appear to have been no studies where numerical bifurcation analysis has been 
applied to spatially extended  models of ice sheets. In this contribution , we make a 
first step in this direction, by looking for multiple equilibria and Hopf bifurcations 
in a  two-dimensional  model of a marine ice sheet. 

\section{Methodology}

\begin{figure}[ht!]\centering
\includegraphics[width=.8\textwidth]{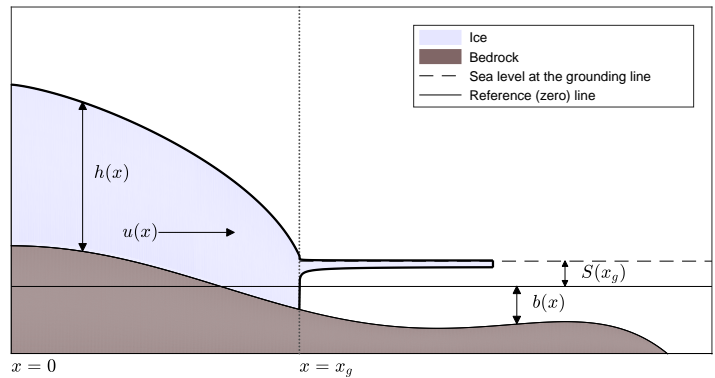}
\caption{\em A two-dimensional marine ice sheet. The ice thickness is given by $h(x)$ 
and the horizontal ice velocity by $u(x)$. Up until the grounding line $x_g$, the ice 
sheet rests on the bedrock $b(x)$. At $x_g$ the ice sheet extends into a floating 
ice shelf. }
\label{fig:schematic}
\end{figure}

Consider in Fig.~\ref{fig:schematic} a two-dimensional marine ice
sheet situated on a bedrock topography in a Cartesian coordinate
system. The ice sheet and bedrock are taken symmetrical, with a
symmetry axis at $x=0$. The grounding line is indicated by $x_g$, the
ice thickness by $h$, and the bedrock by $b$.

\subsection{Model} 

An introduction to ice-sheet and glacier modeling can be found in 
 \cite{Greve_2009} and \cite{Veen_2013}. 
The dynamics of a marine ice sheet is modeled using the shallow-shelf
approximation (SSA), which is obtained by simplifying the full Stokes
problem for gravity driven ice flow \cite{Greve_2009}. The
two-dimensional SSA, as implemented in \cite{Schoof_2007}, is used as
a benchmark problem for the marine ice sheet intercomparison project
(MISMIP, \cite{MISMIP_2012}). Since a thorough comparison with other
results is available, this will be our model of choice for the
bifurcation analysis. Conservation of mass gives:
\begin{equation}
  \D{h}{t}+\D{(uh)}{x} = a, \label{eq:masscons0}
\end{equation}
where $h$ is the ice thickness, $u$ the ice velocity and $a$ the
accumulation rate. On a downward sloping bed
(Fig.~\ref{fig:schematic}), the accumulation $a$ and the ice flux $uh$
at the grounding line are in equilibrium: a positive perturbation of
the grounding line increases both the accumulation and the flux,
leading to a zero net ice growth.

Conservation of momentum gives:
\begin{equation}
  \D{}{x} \left[2A^{-\frac{1}{n}} h
    \Abs{\D{u}{x}}^{\frac{1}{n}-1}\D{u}{x}\right]-
  C\Abs{u}^{(m-1)}u-\rho_i g h \D{(h-b)}{x} = 0, \label{eq:momcons0}
\end{equation} 
where $A$ and $n$ are coefficients of Glen's flow law, a constitutive
relation describing the rheology of ice (\cite{Veen_2013},
\cite{Greve_2009}, typically $n=3$). The ice density is given by
$\rho_i$, $g$ is the gravitational acceleration and $b$ the bedrock
taken positive in the downward direction. The consecutive terms in the
momentum balance \eqref{eq:momcons0} represent longitudinal stresses,
vertical shear stresses, and the driving stress respectively. The
parameters $C$ and $m$ determine the sliding of the ice. Together, $A$
and $C$ effectively describe the size of the transition zone, i.e., the
region in which the grounded sheet becomes afloat and transforms into
an ice shelf.

The left boundary of the domain is located at an \emph{ice divide}, a
location in the ice with zero horizontal flow.  Hence, we take $u=0$ at
$x=0$. As we assume symmetry around $x=0$, we also require that
\begin{equation}
  \D{(h-b)}{x}=0 \quad \text{for}~ x = 0.  \label{eq:leftbdy0}
\end{equation}
We will denote the grounding line by $x_g$. At $x_g$ the ice sheet
becomes afloat and the following flotation condition holds:
\begin{equation}
  \rho_i h = \rho_w b \quad \text{for}~ x = x_g. \label{eq:flotation0}
\end{equation}
From an integration of the shelf flow  \cite{Schoof_2007}, an
extra condition at the grounding line is obtained:
\begin{equation}
  2A^{-\frac{1}{n}} \Abs{\D{u}{x}}^{\frac{1}{n}-1}\D{u}{x} =
  \frac{1}{2}\left(1-\frac{\rho_i}{\rho_w}\right)\rho_i g h \quad
  \text{for}~ x = x_g. \label{eq:shelf0}
\end{equation}
Note that the profile of the shelf is not given by the SSA model, but
obtained using an equilibrium analysis 
\cite{Veen_2013}.

\subsection{Non-dimensional equations}

\newcommand{\transf}[0]{z}

The first difficulty one encounters is the unknown right boundary of the problem given by the grounding line position $x_g$.
As discussed in \cite{Schoof_2007} and \cite{Vieli_2005}, a moving grid approach can be used to track $x_g$.
Using a transformation $\transf = x/x_g$,  the original domain $x \in [0,x_g]$ is mapped 
onto the fixed domain $\transf \in [0,1]$.
As a result, the problem now has three unknowns: $h$, $u$ and $x_g$.
The differential operators are transformed using the chain rule:
\begin{align*}
  \D{}{t} &= \D{}{\tau}\D{\tau}{t} + \D{}{\transf}\D{\transf}{t} = \D{}{\tau} - \frac{\transf}{x_g}\Du{x_g}{t}\D{}{\transf}, \\
  \D{}{x} &= \D{}{\tau}\D{\tau}{x} + \D{}{\transf}\D{\transf}{x} = \frac{1}{x_g}\D{}{\transf},
\end{align*}
where $\transf$ and $\tau$ denote the independent variables in the transformed domain.
Since we only transform in space we have that $\tau = t$.
The transformations of \eqref{eq:masscons0}-\eqref{eq:shelf0} are then given by:
\begin{equation} 
  \D{h}{\tau} - \frac{\transf}{x_g}\Du{x_g}{\tau}\D{h}{\transf}+\frac{1}{x_g}\D{(uh)}{\transf} = a, \label{eq:masscons1}
\end{equation}
\begin{equation}
  \frac{1}{x_g^{\frac{1}{n}+1}}\D{}{\transf} \left[2A^{-\frac{1}{n}} h \Abs{\D{u}{\transf}}^{\frac{1}{n}-1}\D{u}{\transf}\right]- 
  C\Abs{u}^{(m-1)}u-\frac{\rho_i g h}{x_g} \D{(h-b)}{\transf} = 0, \label{eq:momcons1} 
\end{equation}
\begin{equation}
  \D{(h-b)}{\transf}=u=0 \quad \text{for}~ \transf = 0,  \label{eq:leftbdy1}
\end{equation}
\begin{equation}
  \rho_i h = \rho_w b \quad \text{for}~ \transf = 1, \label{eq:flotation1}
\end{equation}
\begin{equation}
  2A^{-\frac{1}{n}} \frac{1}{x_g^{\frac{1}{n}}} \Abs{\D{u}{\transf}}^{\frac{1}{n}-1}\D{u}{\transf} 
  = \frac{1}{2}\left(1-\frac{\rho_i}{\rho_w}\right)\rho_i g h \quad \text{for}~ \transf = 1. \label{eq:shelf1}
\end{equation}

To improve numerical accuracy the equations are non-dimensionalized.
Let
\begin{equation}
h     = h_0       \tilde{h},     \quad b = h_0\tilde{b}, \quad
x_g   = x_0       \tilde{x}_g,   \quad 
\tau  = \tau_0    \tilde{\tau},  \quad 
u     = u_0       \tilde{u},     \quad u_0 = \frac{x_0}{\tau_0},
\end{equation}
with typical thickness $h_0=1 \times 10^3$ m, horizontal extent $x_0=1
\times 10^5$ m and typical timescale $\tau_0 = 100$ y. Substituting
these expressions into  \eqref{eq:masscons1} gives
\begin{equation} 
  \frac{h_0}{\tau_0}\left( \D{\tilde{h}}{\tilde{\tau}} - \frac{\transf}{\tilde{x}_g}\Du{\tilde{x}_g}{\tilde{\tau}}\D{\tilde{h}}{\transf}+\frac{1}{\tilde{x}_g}\D{(\tilde{u}\tilde{h})}{\transf}\right) = a \Leftrightarrow  \D{\tilde{h}}{\tilde{\tau}} - \frac{\transf}{\tilde{x}_g}\Du{\tilde{x}_g}{\tilde{\tau}}\D{\tilde{h}}{\transf}+\frac{1}{\tilde{x}_g}\D{(\tilde{u}\tilde{h})}{\transf} = \Omega,   \label{eq:masscons2}
\end{equation}
where we let $\Omega = \frac{\tau_0}{h_0} {a}$. Similarly, the non-dimensionalized version of \eqref{eq:momcons1} is given by
\begin{equation}
  \frac{1}{\tilde{x}_g^{\frac{1}{n}+1}}\D{}{\transf} \left[\tilde{h} \Abs{\D{\tilde{u}}{\transf}}^{\frac{1}{n}-1}\D{\tilde{u}}{\transf}\right]- {\Gamma}\Abs{\tilde{u}}^{(m-1)}\tilde{u}-{\Lambda}\frac{\rho_i g \tilde{h}}{\tilde{x}_g} \D{(\tilde{h}-\tilde{b})}{\transf} = 0, \label{eq:momcons2} 
\end{equation}
where we introduce the new constants 
\begin{equation}\Gamma := C\left(\frac{x_0}{\tau_0}\right)^m \Big/ \left(2A^{-\frac{1}{n}}\left(\frac{x_0}{\tau_0}\right)^{\frac{1}{n}} \frac{h_0}{x_0^{\frac{1}{n}+1}}\right) \quad \text{and} \quad \Lambda := \frac{h_0^2}{x_0} \Big/ \left(2A^{-\frac{1}{n}}\left(\frac{x_0}{\tau_0}\right)^{\frac{1}{n}} \frac{h_0}{x_0^{\frac{1}{n}+1}}\right).\end{equation}
Finally, at the boundaries we obtain
\begin{equation}
  \D{(\tilde{h}-\tilde{b})}{\transf}=\tilde{u}=0 \quad \text{for}~ \transf = 0,  \label{eq:leftbdy2}
\end{equation}
\begin{equation}
  \rho_i \tilde{h} = \rho_w \tilde{b} \quad \text{for}~ \transf = 1, \label{eq:flotation2}
\end{equation}
\begin{equation}
  \frac{1}{\tilde{x}_g^{\frac{1}{n}}} \Abs{\D{\tilde{u}}{\transf}}^{\frac{1}{n}-1}\D{\tilde{u}}{\transf} 
  = \Sigma\left(1-\frac{\rho_i}{\rho_w}\right)\rho_i g \tilde{h} \quad \text{for}~ \transf = 1, \label{eq:shelf2}
\end{equation}
with
\begin{equation}\Sigma := \frac{1}{2} h_0 \Big/ \left(2A^{-\frac{1}{n}}\frac{1}{\tau_0^{\frac{1}{n}}}\right).\end{equation}

\subsection{Numerical implementation}\label{sec:implementation}

From here on we omit the tildes and assume the unknowns are non-dimensional.
As in \cite{Schoof_2007}, the domain $\transf \in [0,1]$ is discretized using a staggered grid with a fixed mesh-width: $\Delta \transf = {1}/{(N-1/2)}$.
The left boundary is taken at the vertex $i=1$ and the right boundary at the cell center $i=N+1/2$.
Thus, for $i=1,2,\ldots,N$, we have vertices at $\transf_i = \Delta \transf (i-1)$ and cell centers at $\transf_{i+1/2} = \Delta \transf (i-1/2)$.
The discretized solution values for ice thickness are located at the vertices $h_i$, while the values for ice velocity are positioned at the cell centers $u_{i+1/2}$.

The transformed and non-dimensionalized continuity equation \eqref{eq:masscons2} is discretized using a central difference for the stretching and an upwind discretization for the flux:
\begin{normalsize}\begin{equation}
  \Du{h_i}{\tau} - \frac{\transf_i}{x_g} \left(\frac{h_{i+1}-h_{i-1}}{2\Delta\transf}\right)\Du{x_g}{\tau} =
                 - \frac{h_i\left(u_{i+1/2}+u_{i-1/2}\right)-h_{i-1}\left(u_{i-1/2}+u_{i-3/2}\right)}{2 x_g \Delta \transf} + \Omega_i. \label{eq:massconsdisc}
\end{equation}\end{normalsize}
At the left boundary symmetry requires
\begin{normalsize}\begin{align*}
{(h_2-h_0)}/{(2\Delta \transf)} &= 0 ~~\text{(central)}, \\ {(h_1-h_0)}/{\Delta \transf} &= 0 ~~\text{(upwind)}, \\ u_{3/2}+u_{1/2} &= 0, \quad u_{5/2}+u_{-1/2} = 0.
\end{align*}\end{normalsize}
Using these expressions we can resolve the dependence on nonexistent grid-points. For $i=1$ and $i=2$, mass conservation is therefore given by
\begin{normalsize}\begin{align}
  (i=1) \qquad& \Du{h_1}{\tau} = - \frac{h_{1}\left(u_{3/2}+u_{5/2}\right)}{2 x_g \Delta \transf} + \Omega_1, \label{eq:massconslbdy1}\\
 (i=2) \qquad&  \Du{h_2}{\tau} - \frac{\transf_2}{x_g} \left(\frac{h_{3}-h_{1}}{2\Delta\transf}\right)\Du{x_g}{\tau} = - \frac{h_2\left(u_{5/2}+u_{3/2}\right)}{2 x_g \Delta \transf} + \Omega_2. \label{eq:massconslbdy2}
\end{align}\end{normalsize}
Note that at the rightmost vertex ($i=N$), the right hand side of the discretized mass conservation \eqref{eq:massconsdisc} does not contain any dependencies on nonexistent grid-points.
In the left hand side we will need to use a one-sided difference for the stretching term.

Define $\Delta u_i := u_{i+1/2}-u_{i-1/2}$.
The momentum conservation \eqref{eq:momcons2} is discretized using central differences:
\begin{normalsize}\begin{align}
0 ~=~ & \frac{1}{(x_g\Delta\transf)^{1+1/n}}  \left[h_{i+1}\Abs{\Delta u_{i+1}}^{1/n-1}\Delta u_{i+1}-h_i\Abs{\Delta u_{i}}^{1/n-1}\Delta u_{i}\right] & \nonumber\\
&-~ \Gamma\Abs{u_{i+1/2}}^{m-1}u_{i+1/2} - \Lambda \left(\frac{h_i+h_{i+1}}{2}\right)\frac{\rho_i g}{x_g\Delta\transf}\left[h_{i+1}-b_{i+1}-h_i+b_i\right]. \label{eq:momconsdisc}
\end{align}\end{normalsize}
At the left boundary we let $\Delta u_1 = 2 u_{3/2}$. 
At the right boundary we impose the following discretization of \eqref{eq:shelf2} with a substituted flotation condition $\rho_i h_N = \rho_w b_N$ (cf.~\eqref{eq:flotation2}): 
\begin{equation}
0 = \frac{1}{(x_g\Delta\transf)^{1/n}}\Abs{\Delta u_N}^{1/n-1}\left(\Delta u_N\right)-\Sigma\left(1-\frac{\rho_i}{\rho_w}\right){\rho_w g b_N}. \label{eq:shelfdisc} 
\end{equation}

The discretization contains $N$ unknown $h_i$, $N$ unknown $u_{i+1/2}$
and an unknown grounding line position $x_g$. To achieve a closed
system of $2N+1$ equations, the flotation criterion at the cell center
$\transf_{N+1/2}$ is prescribed using an extrapolation of the
thickness, which gives the closing requirement:
\begin{equation}
0 = 3h_N-h_{N-1}-2\frac{\rho_w}{\rho_i}b_{N}.
\end{equation}
Finally we obtain a problem of the form 
\begin{equation}
  M \Du{}{t} \bvec{x} = F(\bvec{x}, \lambda), ~\text{with}~ \bvec{x} = \begin{bmatrix} \bvec{h} \\ \bvec{u} \\ x_g \end{bmatrix}.
\end{equation}
The unknown functions are discretized: $\bvec{h},\bvec{u} \in
\mathbb{R}^N $. The real-valued matrix $M \in
\mathbb{R}^{(2N+1)\times(2N+1)}$ and nonlinear operator
$F:\mathbb{R}^{2N+1} \to \mathbb{R}^{2N+1}$ are given by
\begin{equation}
  M = \begin{bmatrix}~~I~ & 0 & M_{\text{str}}(\bvec{h},x_g) \\ 0 & 0 & 0 \\ 0 & 0 & 0 \end{bmatrix}, \quad F = \begin{bmatrix} F_{\text{mass}}(\bvec{h},\bvec{u},x_g) \\ F_{\text{mom}}(\bvec{h},\bvec{u},x_g) \\ F_{\text{flot}}(\bvec{h},x_g)\end{bmatrix}.
\end{equation}
Here, $M_{\text{str}}(\bvec{h},x_g) \in \mathbb{R}^{N}$ is the
discretization of the stretching in the left-hand side of Equations
\eqref{eq:massconsdisc}-\eqref{eq:massconslbdy2}.
$F_{\text{mass}}(\bvec{h},\bvec{u},x_g) \in \mathbb{R}^{N}$ is given
by the right-hand side of discretizations
\eqref{eq:massconsdisc}-\eqref{eq:massconslbdy2}. Similarly,
$F_{\text{mom}}(\bvec{h},\bvec{u},x_g) \in \mathbb{R}^{N}$ and
$F_{\text{flot}}(\bvec{h},\bvec{u},x_g) \in \mathbb{R}$ are given by
the right hand sides of the discretized momentum equation and
flotation criterion \eqref{eq:momconsdisc}-\eqref{eq:shelfdisc}.

\subsection{Pseudo-arclength continuation}\label{sec:pseudoarc}

The discretized equations give a problem of the form
\begin{equation}
  M \Du{\bvec{x}}{t}  = F(\bvec{x}, \lambda),
\end{equation}
where $M$ and $F(\cdot)$ are linear and non-linear operators
respectively. We explicitly introduce the parameter dependence
$\lambda$ since we are interested in solution branches
$(\bvec{x},\lambda)$ satisfying $F(\bvec{x}, \lambda) = 0$. For
example, our first parameter of interest will be the temperature,
which is present in the coefficient $A$ in Glen's flow law.

Various continuation techniques exist to trace a stationary solution
branch while varying a parameter. A successful approach is to
parameterize a solution branch with a pseudo-arclength parameter $s$:
$\gamma(s) = (\bvec{x}(s),\lambda(s))$ and impose an approximate
normalization condition on the tangent, to close the system of
equations: $\dot{\bvec{x}}^T(\bvec{x}-\bvec{x}_0) +
\dot{\lambda}(\lambda-\lambda_0) - \Delta s = 0$, where
$(\bvec{x}_0,\lambda_0)$ is an initial known stationary solution,
$(\dot{\bvec{x}},\dot{\lambda})$ the tangents w.r.t.~the arclength
parameter at $(\bvec{x}_0,\lambda_0)$ and $\Delta s$ a specified step
size \cite{Dijkstra_2005, Kuznetsov_2004}.

To find a new point on the solution branch a predictor-corrector method is used. 
A suitable tangent predictor is given by 
\begin{align*}
\bvec{x}^1 = \bvec{x}_0 + \Delta s~ \dot{\bvec{x}}, \\ 
\bvec{\lambda}^1 = \bvec{\lambda}_0 + \Delta s~ \dot{\lambda}.
\end{align*}
Note that the prediction is denoted by $(\bvec{x}^1,\lambda^1)$, whereas an actual new 
solution will be denoted by $(\bvec{x}_1,\lambda_1)$. The correction onto the solution 
branch is made through the solution of the nonlinear system given by
\begin{align}
F(\bvec{x}, \lambda) = 0, \\ 
\dot{\bvec{x}}^T(\bvec{x}-\bvec{x}_0) + \dot{\lambda}(\lambda-\lambda_0) - \Delta s = 0.
\end{align}
A Newton-Raphson root finding procedure, initialized with the prediction 
$(\bvec{x}^1,\lambda^1)$, gives the following iteration:
\begin{equation}
\begin{bmatrix}
F_\bvec{x}         &  F_\lambda \\
\dot{\bvec{x}}^T  &  \dot{\lambda}
\end{bmatrix} 
\begin{bmatrix} \Delta \bvec{x} \\ \Delta \lambda
\end{bmatrix}
= \begin{bmatrix}
- F(\bvec{x}^k, \lambda) \\
\Delta s - \dot{\bvec{x}}^T(\bvec{x}^k -\bvec{x}_0) - \dot{\lambda}(\lambda^k-\lambda_0) 
\end{bmatrix}, \label{eq:PSEUDOARCNR}
\end{equation}
where $\Delta \bvec{x} := \bvec{x}^{k+1}- \bvec{x}^{k}$, $\Delta
\lambda := \lambda^{k+1} - \lambda^k$ and $[F_\bvec{x}, F_\lambda]$ is
the Jacobian matrix of $F$. If this iteration converges a new
stationary solution $(\bvec{x}_1,\lambda_1)$ has been found. At a fold
bifurcation the Jacobian matrix $F_\bvec{x}$ will have a zero
eigenvalue, yet the system in \eqref{eq:PSEUDOARCNR} remains
non-singular,  and the continuation is able to trace the solution branch
into its unstable domain.

When a stationary solution $\bar{\bvec{x}}$, satisfying
\begin{equation}
F(\bar{\bvec{x}}, \lambda) = 0,
\end{equation}
has been found, its stability can be investigated with a perturbation $\bar{\bvec{x}} + \tilde{\bvec{x}}$ and a Taylor expansion around the stationary solution: 
\begin{align}
 M \Du{}{t}\left( \bar{\bvec{x}} + \tilde{\bvec{x}}\right) = M \Du{}{t}\bar{\bvec{x}} + M \Du{}{t}\tilde{\bvec{x}} &= F(\bar{\bvec{x}} + \tilde{\bvec{x}}, \lambda) \approx F(\bar{\bvec{x}}, \lambda) +  F_{\bar{\bvec{x}}}(\bar{\bvec{x}}, \lambda)\tilde{\bvec{x}} ~\Leftrightarrow~ \nonumber\\
M \Du{}{t}\tilde{\bvec{x}} = F_{\bar{\bvec{x}}}(\bar{\bvec{x}}, \lambda)\tilde{\bvec{x}}. \label{eq:perturbed}
\end{align}
Solutions of \eqref{eq:perturbed} are of the form $\tilde{\bvec{x}} = \hat{\bvec{x}}\Exp{\sigma t}$. Substitution gives a generalized eigenvalue problem:
\begin{equation} 
  \sigma  M \hat{\bvec{x}} = F_{\bar{\bvec{x}}}(\bar{\bvec{x}}, \lambda) \hat{\bvec{x}}.
  \label{eq:eigv}
\end{equation}
The stability of a stationary solution depends on the sign of the real
part of the eigenvalues. If we find an eigenvalue with positive real
part the solution contains a growing mode and is thus unstable.

\section{Results} \label{sec:results}

\subsection{Bifurcation diagram} 

First we consider the case of constant accumulation, represented by a
constant $a$, together with a fixed bedrock $b(x)$. The bedrock
profile is chosen as in \cite{Schoof_2007}:
\begin{equation}
 b(x) = -\left(729 - 2184.8\left(\frac{x}{S}\right)^2 +
 1031.72\left(\frac{x}{S}\right)^4
 -151.72\left(\frac{x}{S}\right)^6\right), \label{eq:initbedrock}
\end{equation} 
with $S = 750\times 10^3 \textup{m}$. To find a stationary solution,
an ice sheet surface profile of the form
\begin{equation}
  s = h-b =~ C_1 \sqrt{1-\left(\frac{x}{x_g}\right)^2}+{C_2},
  ~\text{with}~ x \in [0,x_g],
\end{equation}
is used as initial guess for the Newton-Raphson iteration solving
$F(\bvec{x}, \lambda) = 0$. Typically, $C_1 = 3\times 10^3$ and $C_2$
is chosen such that the flotation criterion is satisfied:
$C_2=({\rho_w}/{\rho_i}-1)b_g$, where $b_g$ is the bedrock at the
grounding line. The analytic profile contains a steep gradient at the
grounding line, which is essential for a good convergence of the root
finding process.

\begin{figure}[t!]
\includegraphics[width=.405\textwidth]{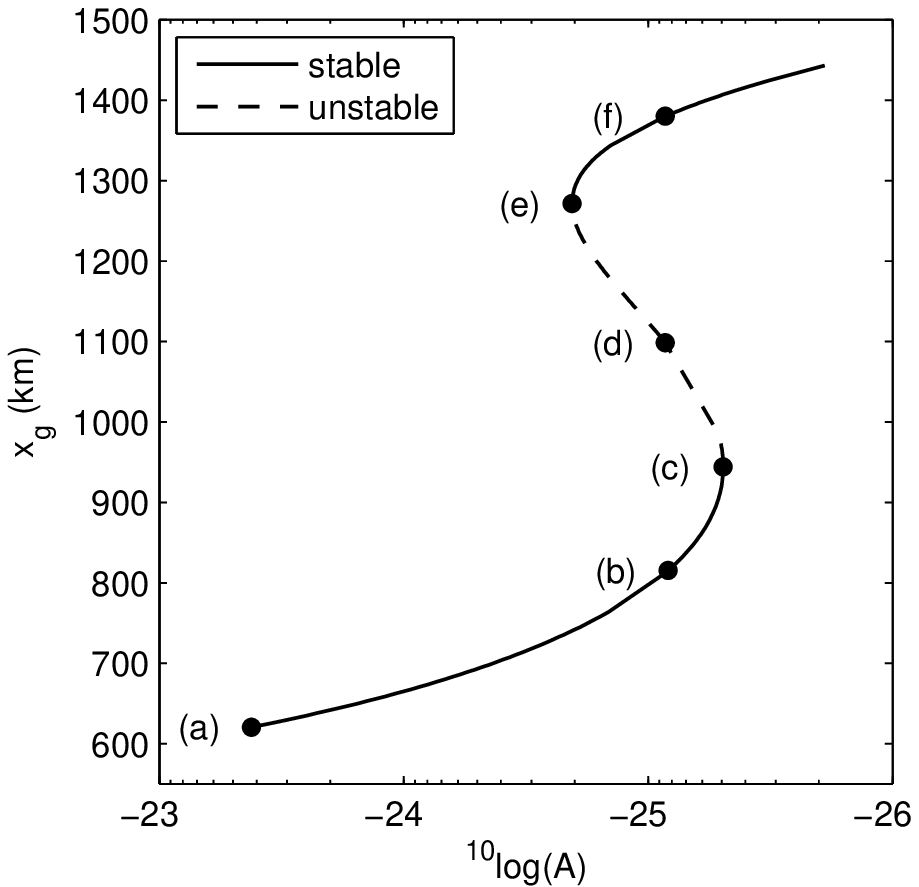}\quad
\includegraphics[width=.595\textwidth]{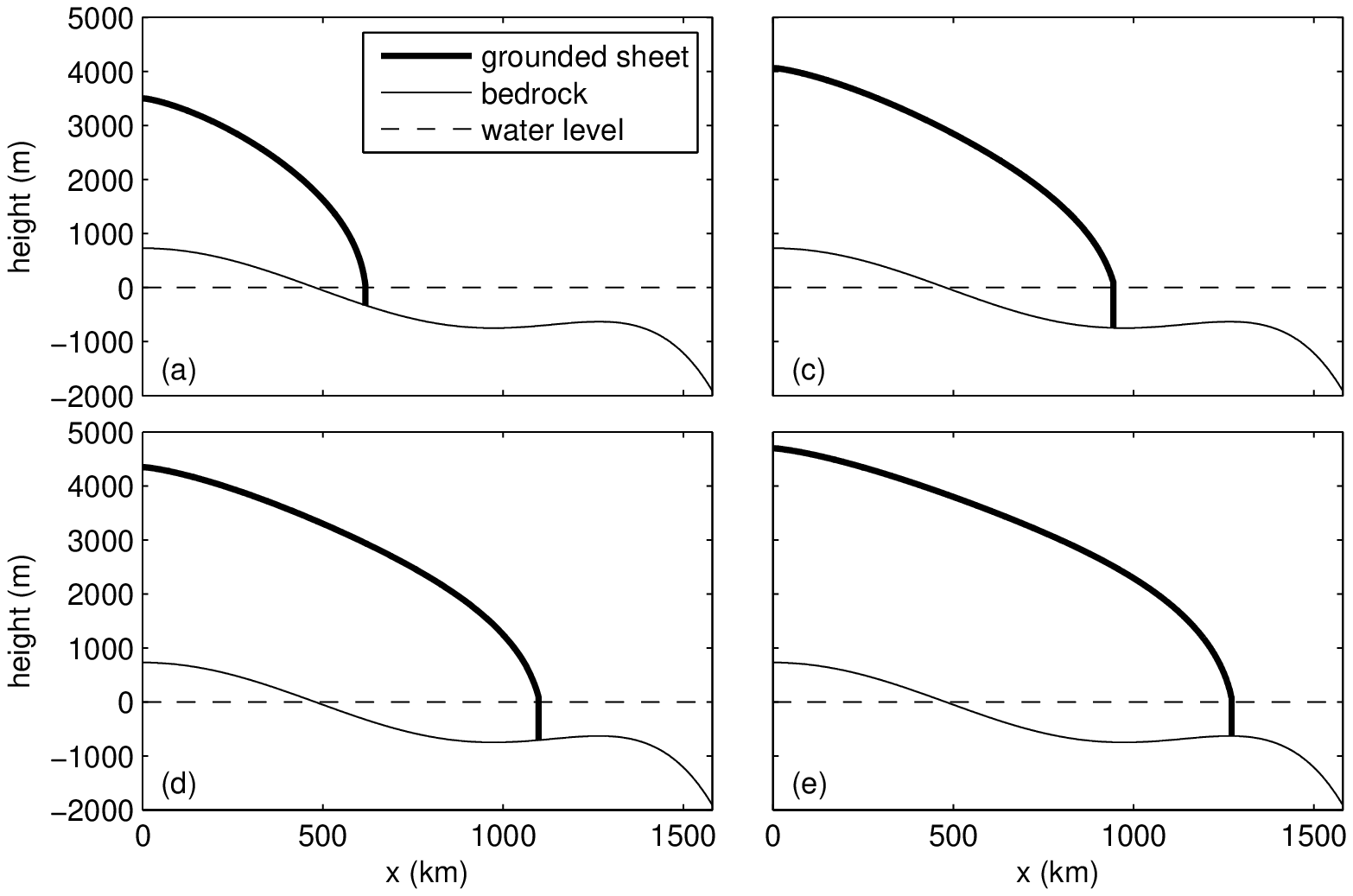}
\caption{One-parameter bifurcation diagram (left) and solutions (right), decreasing the parameter $A$ in the SSA model which corresponds to a decrease in temperature and an increase in ice growth. The bedrock contains an upward slope which admits multiple steady states  ((b),(d) and (f)) for a constant parameter. Eigenvalue analysis  shows two saddle-node bifurcations: a single eigenvalue crosses the imaginary axis to the positive right half plane at (c) and returns to the left half plane at (e). The number of grid-points is $N=1600$; the other parameters are given in Table \ref{tab:rempars}.}
\label{fig:results0}
\end{figure}

After an initial equilibrium profile is obtained for $A = 4.6416\times
10^{-24}$, a pseudo-arclength continuation traces the solution branch
$\gamma(s)$ in the direction of decreasing $A$, see Figure
\ref{fig:results0}. The presented bifurcation diagram confirms the
multiple equilibria regime associated with the hysteretic behavior
shown in \cite{Schoof_2007}. For a fixed value of the parameter $A$, 
three equilibria are distinguished and marked as (b), (d) and (f). At
the point (c) a saddle-node bifurcation occurs and an eigenvalue is
observed to cross the imaginary axis to the right half plane. At (e),
the same eigenvalue returns to the right half plane through a second
saddle-node bifurcation. The values of the parameters are summarized
in Table \ref{tab:rempars}.

\begin{table}[htbp]\centering
  \begin{footnotesize}
  \begin{tabular}{|c|c|} 
    \hline \vspace{-2.5ex} &\\
 $C$ & $7.624 \times 10^6~~\text{Pa}~\text{m}^{-1/3}~\text{s}^{1/3}$\\
 $m$ & $1/3$               \\
 $n$ & $3$                 \\
 $g$ & $9.8$~~$\text{m}$~$\text{s}^{-2}$               \\
 $\rho_i$ & $900$~~$\text{kg}~\text{m}^{-3}$         \\
 $\rho_w$ & $1000$~~$\text{kg}~\text{m}^{-3}$        \\
 $a$      & $0.3$~~$\text{m}~\text{y}^{-1}$     \\\hline     
  \end{tabular} \quad
\begin{tabular}{|c|c|} 
    \hline \vspace{-2.5ex} &\\
      &  $A$ ($\text{s}^{-1}~\text{Pa}^{-3}$) \\    \hline \vspace{-2.2ex} &\\
  (a) & $4.6416\times 10^{-24}$ \\
  (b) & $8.5014\times 10^{-26}$ \\
  (c) & $4.9274 \times 10^{-26}$\\
  (d) & $8.5014\times 10^{-26}$ \\
  (e) & $2.0450 \times 10^{-25}$\\
  (f) & $8.5014\times 10^{-26}$ \\ \hline
  \end{tabular}\end{footnotesize}
  \caption{Parameter values for the experiment in Figure \ref{fig:results0}, similar to the values chosen in \cite{Schoof_2007} and \cite{MISMIP_2012}.}
  \label{tab:rempars}
\end{table}

\subsection{Numerical accuracy}
In order to investigate the numerical accuracy of the discretization
and continuation methodology we perform a convergence experiment, see
Table \ref{tab:convres}. Let the error in the approximated bifurcation
point $A_N$ be proportional to a power of the mesh-width:
\begin{equation}
A_N = A + \alpha (\Delta \transf)^\beta,
\end{equation}
where $A$ denotes the actual bifurcation point and $\alpha$ and
$\beta$ are constants. We define a difference between subsequent
mesh-halvings $D_{N}:=A_{N}-A_{N/2}$ and let $\Delta \transf \approx
1/N$. Then, the ratio between consecutive differences only depends on
the power $\beta$:
\begin{equation}
R_{N} := \frac{D_{N/2}}{D_{N}} = \frac{A_{N/2}-A_{N/4}}{A_{N}-A_{N/2}}=\frac{(1/2)^\beta-1}{(1/4)^\beta-(1/2)^\beta}.
\end{equation} 
\begin{table}[ht!]
  \centering
  \begin{tabular}{|c|c|c|c|}
    \hline
   $N$        & $A_N$   &  $D_N$  & $R_N$    \\ \hline  \vspace{-2ex} & & & \\
$  50 $  &  $2.61941\times 10^{-26} $ &  & \\
$ 100 $  &  $3.48732\times 10^{-26} $ &  $8.67909\times 10^{-27}$  & \\
$ 200 $  &  $4.14983\times 10^{-26} $ &  $6.62512\times 10^{-27}$  & $1.31003$  \\
$ 400 $  &  $4.56744\times 10^{-26} $ &  $4.17606\times 10^{-27}$  & $1.58645$  \\
$ 800 $  &  $4.80250\times 10^{-26} $ &  $2.35065\times 10^{-27}$  & $1.77655$  \\
$1600 $  &  $4.92741\times 10^{-26} $ &  $1.24911\times 10^{-27}$  & $1.88186$  \\\hline
  \end{tabular}
  \caption{Convergence of the first bifurcation (point (c) in Figure \ref{fig:results0}).}\label{tab:convres}
\end{table}

From Table \ref{tab:convres} we suspect $R_N \to 2$ as $N$ becomes
large, corresponding to $\beta = 1$. The scheme must therefore be of
first-order accuracy, which is undoubtedly due to the first-order
upwind discretization in the continuity equation
\eqref{eq:massconsdisc}. Unfortunately, the upwind discretization of
the ice flux is essential to the stability of the scheme. We conclude
that a higher order upwind scheme should be considered in
\eqref{eq:massconsdisc}.

\subsection{Mechanism} 

The advantage of the approach chosen here is that the spatial patterns
of perturbations destabilizing  the marine ice sheet can be determined
from the eigenvectors in \eqref{eq:eigv}.
\begin{figure}\centering
  \includegraphics[width=\textwidth]{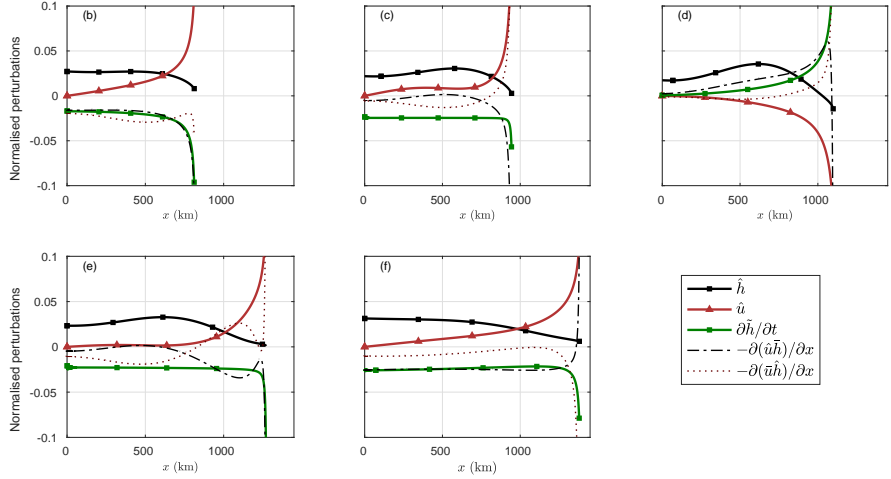}
  \caption{Normalized components of the eigenmode that becomes
    unstable. The corresponding steady states are located at the
    points (b)-(f), described in \figref{results0} and
    \tabref{rempars}.  We distinguish between a perturbation
    pattern related to sheet thickness $\hat{h}$ and a pattern related
    to ice velocity $\hat{u}$. The signs of the eigenvectors are taken
    such that the perturbation in the grounding line is positive. From
    the eigenvectors and the equilibrium solution we compute a
    normalized spatial pattern of the evolution $\partial \tilde{h} /
    \partial t$, together with normalized components $ -\partial
    (\hat{u} \bar{h}) / \partial x$ and $ -\partial (\bar{u} \hat{h})
    / \partial x$.}
  \label{fig:eigenvectors}
\end{figure}
For the unstable equilibrium (d) in \figref{results0} it is of
interest to examine the eigenmode with a positive growth factor,
showing in detail the characteristics of the instability. The
eigenvector is made available using \eqref{eq:eigv} and is depicted
for the steady states (b),(c),(d),(e) and (f) in
\figref{eigenvectors}. The perturbations in thickness and velocity are
taken corresponding to a positive grounding line perturbation. Note
that a perturbation of the solution vector has the form
$\hat{\bvec{x}}e^{\sigma 
  t}=[\hat{\bvec{h}},\hat{\bvec{u}},\hat{x}_g]^Te^{\sigma t}$, with
$\hat{x}_g$  the scalar grounding line perturbation. An eigenvector with corresponding
eigenvalue $\sigma > 0$ and $\hat{x}_g < 0$ gives the destabilizing 
pattern for unstable ice sheet retreat. By adjusting the sign of the
eigenvector, such that $\hat{x}_g > 0 $, we restrict our exposition to
destabilizing  patterns for unstable ice sheet growth. 

At the grounding line $x_g$, the perturbation  of the unstable
steady state (d)  shows a slight decrease in ice thickness, while
at (b),(c),(e) and (f) a slight increase is observed. In the interior
of the ice sheet a relatively large increase in ice thickness is
visible for the unstable equilibrium (d), indicating interior ice
growth due to an imbalance between global accumulation and ice flux at
the grounding line.

The velocity perturbation at the grounding line shows an increase
for stable states and a clear decrease for the unstable state (d).
Together with the negative perturbation in thickness this implies
that, at (d), there must be a decrease in flux $uh$ across the
grounding line for a positive perturbation $\hat{x}_g > 0$. An
increase in grounding line position implies a rise in global
accumulation, hence the reduction in grounding line flux implies a net
ice growth, confirming the marine ice sheet instability hypothesis.
Note that a similar result holds if we take the perturbation in
$x_g$ negative, giving a net ice loss and a retreat from equilibrium.

The continuation approach allows an efficient computation of flux
perturbations using \eqref{eq:eigv}. From a linear stability analysis
of \eqref{eq:masscons0} with perturbation $\tilde{h} = c \hat{h},
~\tilde{u} = c \hat{u}$ around an equilibrium $\bar{h},\bar{u}$ we
obtain an evolution equation for the  thickness perturbation
$\tilde{h}$:
\begin{align}
  \D{\tilde{h}}{t} + c \D{\hat{q}}{x} = 0, & \quad \text{with}~ c > 0, \\
  \hat{q}(x) = \hat{u}\bar{h} + \bar{u}\hat{h}, \label{eq:fluxpert}&
\end{align}
where we neglect higher order terms. At the unstable steady state (d)
in \figref{eigenvectors}, the thickness perturbation (green
squares) shows positive growth, whereas
the other points show a dampening. These patterns are determined by
spatial derivatives of the perturbed advection of the steady thickness
$\hat{u}\bar{h}$ (black dash-dotted line)and the advected thickness perturbation
$\bar{u}\hat{h}$ (red dotted line).  The latter clearly dominates the instability in (d).
Note that at the bifurcation points (c) and (e) the components of the
perturbation flux have a compensating effect.

To investigate how a perturbation changes from stable to unstable
through the saddle-node bifurcation $L_1$, we compute the accumulation
and grounding line fluxes. The steady state $(\bar{h},\bar{u})$ gives
a balance:
\begin{equation}
  \bar{q}(x) = \bar{u}\bar{h} = ax. \label{eq:steadymasscons0}
\end{equation}
\begin{figure}\centering

  \includegraphics[height=.30\textwidth]{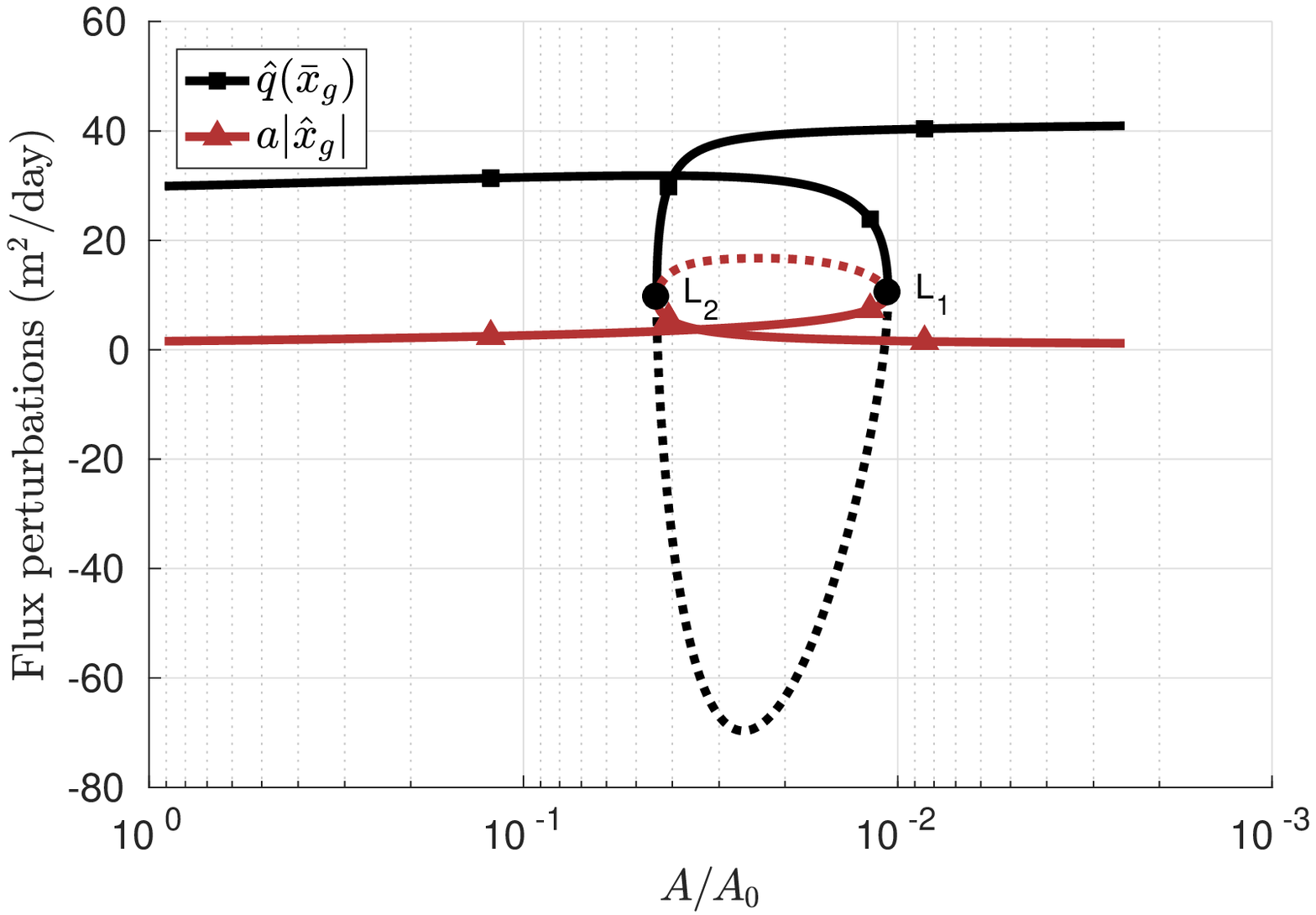} {\footnotesize(a)} 
  \includegraphics[height=.30\textwidth]{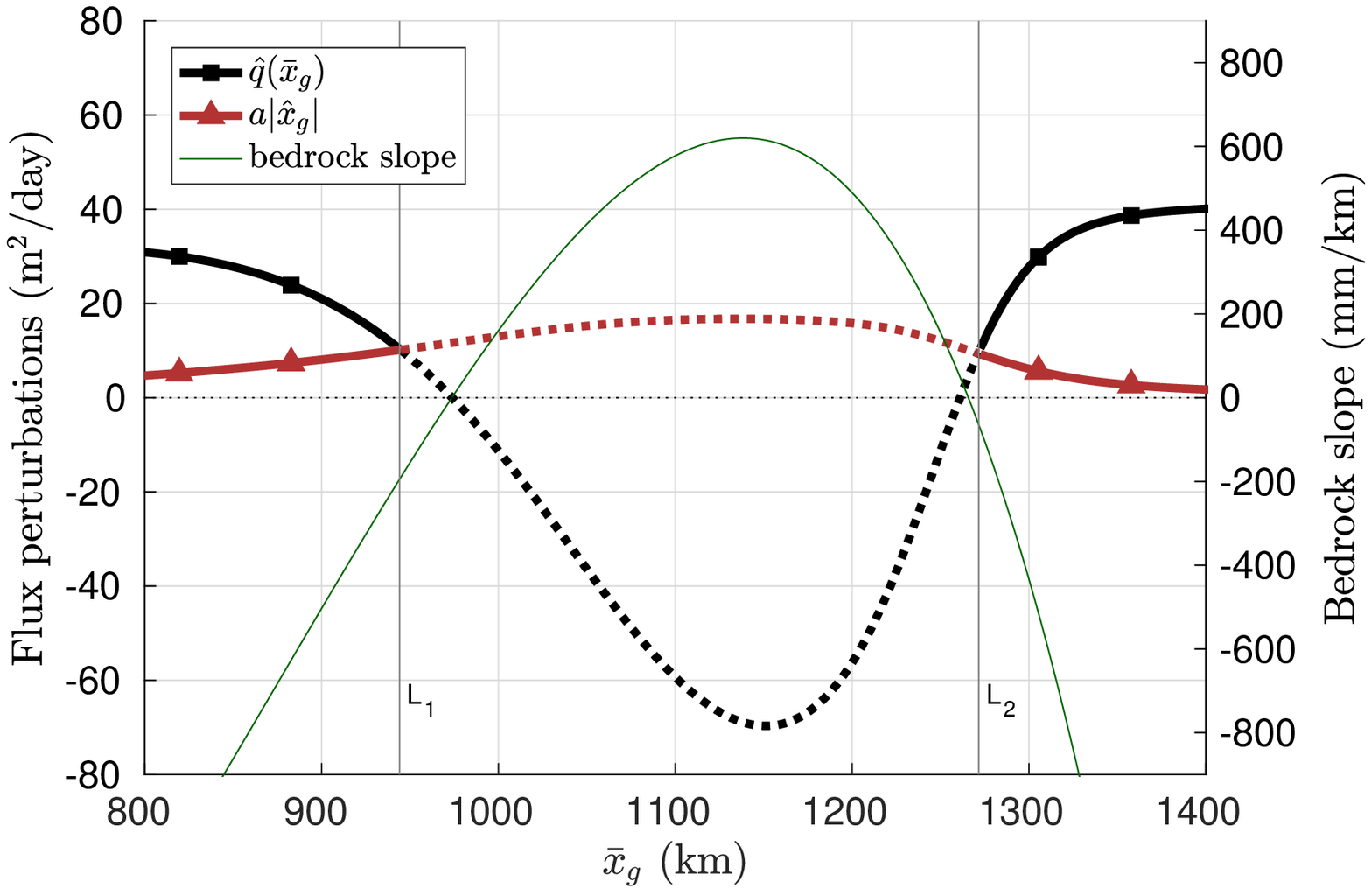}   {\footnotesize(b)} 

  \caption{Perturbations in grounding line and accumulation flux
    against the parameter $A$ (a) and as a function of $x_g$ (b),
    together with the bedrock slope. The first and second saddle-node
    bifurcations are marked $L_1$ and $L_2$. Dashed lines correspond
    to growing perturbations of unstable steady states. The
    perturbations $\hat{q}(\bar{x}_g)$ and $a\Abs{\hat{x}_g}$
    correspond to a positive grounding line perturbation $\hat{x}_g >
    0$. At $L_1$ and $L_2$, the grounding line bedrock slope is $-195$
    mm/km and $-65$ mm/km respectively. The maximum slope in the
    unstable regime is 620 mm/km. Note that grounding line flux
    perturbations depend on the steady grounding line position
    $\bar{x}_g$, whereas accumulation flux perturbations depend on the
    grounding line perturbation $\hat{x}_g$.}
  \label{fig:fluxandaccum}
\end{figure}
In \figref{fluxandaccum} we show perturbations of the balance
\eqref{eq:steadymasscons0} at the grounding line. A perturbation in
accumulation flux is given by $a\Abs{\hat{x}_g}$, a grounding line
flux perturbation by $\hat{q}(\bar{x}_g)$ in \eqref{eq:fluxpert}. The
perturbations are plotted against $A/A_0$ (\figref{fluxandaccum}a) and
$\bar{x}_g$ (\figref{fluxandaccum}b), together with the bifurcation
points and the bedrock slope. At the first saddle-node bifurcation
$L_1$, the flux $\hat{q}(x_g)$ becomes smaller than the accumulation
$a\Abs{\hat{x}_g}$. Beyond this point, a change in accumulation due to
$\hat{x}_g$ is not balanced by the grounding line flux and, hence, the
perturbation $\tilde{\bf{x}}$ changes from damped to growing. At
$L_2$, $\hat{q}(x_g)$ becomes greater than $a\Abs{\hat{x}_g}$ and the
mode is damped again.

In \figref{fluxandaccum}b we also plot the bedrock slope, taken
positive when the bed elevation increases with $x$, that is
\begin{equation}
  r_\text{bed} = - b^\prime(x_g),
\end{equation}
with $b(x)$ as in \eqref{eq:initbedrock}. Note that the sign switch in
$\hat{q}(x_g)$ coincides with the sign switch in the bedrock slope.
The grounding line flux will increase for a positive $\hat{x}_g$
between the bifurcation $L_1$ and the point of zero bedrock slope,
but, since the change is less than the change in accumulation, the ice
sheet will grow. Thus, \figref{fluxandaccum} confirms that an
eigenvalue of the `full model' in \cite{Schoof_2007} becomes positive
when $a\Abs{\hat{x}_g} - \hat{q}(x_g) > 0$.

Using a continuation of steady states with the original SSA equations,
we find that the flux perturbations and their relative magnitude fully
describe the instability mechanism, confirming the analysis in
\cite{Schoof_2012}. In addition, the eigenvectors reveal destabilizing 
interior patterns with, most interestingly, interior thickness
anomalies and their advection. These turn out to play a major role in
the unstable growth and retreat of the ice sheet.

\subsection{Glacial isostatic adjustment}\label{sec:extentions}

The simplest model describing the interaction between an ice sheet and
the underlying bedrock is a local lithosphere, relaxing asthenosphere
(LLRA) model. An equilibrium argument \cite{Greve_2009} gives:
\begin{equation}\rho_a g (b^*-b^0) = \rho_i g h ~\Rightarrow~ (b^*-b^0) = \frac{\rho_i}{\rho_a} h,\end{equation}
with $\rho_a$ the density of the asthenosphere, $b^*$ the equilibrium bedrock and 
$b^0$ the initial, load-free bedrock. Due to the highly viscous asthenosphere the 
equilibrium bedrock is reached after a significant response time $\tau_a$. The 
evolution of the bedrock can be modeled using 
\begin{equation}
\D{b}{t} = \frac{1}{\tau_a}\left[(b^*-b^0)-(b-b^0) \right] = \frac{1}{\tau_a}\left[\frac{\rho_i}{\rho_a}h-(b-b^0) \right], \label{eq:bedrock0}
\end{equation}
with $b$ the actual bedrock profile.

Adding this elastic bedrock to the SSA model means introducing a new unknown and a new 
discretized differential equation. To remain consistent with the implementation in Section 
\ref{sec:implementation}, we need to perform the transformation $\transf = x/x_g$, giving
\begin{equation}
\D{b}{\tau} - \frac{\transf}{x_g}\Du{x_g}{\tau}\D{b}{\transf}  =  \frac{1}{\tau_a}\left[\frac{\rho_i}{\rho_a}h-(b-b^0) \right]. \label{eq:bedrock1}
\end{equation}
Non-dimensionalizing \eqref{eq:bedrock1} is straightforward. Using a central difference for the 
stretching we obtain the following discretization:
\begin{equation}
\Du{b_i}{\tau} - \frac{\transf}{x_g}\Du{x_g}{\tau}\left(\frac{b_{i+1}-b_{i-1}}{2\Delta \transf}\right)  =  \frac{1}{\tau_a}\left[\frac{\rho_i}{\rho_a}h_i-(b_i-b^0_i) \right]. \label{eq:bedrock2}
\end{equation} 
Symmetry at the left boundary gives a vanishing spatial derivative. At
the grounding line the pressure exerted by a column of ice equals that
of the column of water: $\rho_i h_N = \rho_w b_N$. Substituting the
flotation criterion gives the following discretization for the right
boundary of the bedrock equation:
\begin{equation}
\Du{b_N}{\tau} - \frac{\transf}{x_g}\Du{x_g}{\tau}\left(\frac{\frac{\rho_i}{\rho_w}h_N-b_{N-1}}{\Delta \transf}\right)  =  \frac{\rho_w - \rho_a}{\rho_a\tau_a}\left[b_N-\frac{\rho_a}{\rho_a-\rho_w}b^0_N\right]. \label{eq:bedrockbdy}
\end{equation} 
Equation \eqref{eq:bedrockbdy} acts as a Dirichlet boundary condition
in the stationary case. Note that, with this boundary, we assume the
bedrock is adjusted regardless of the type of material that is on top
of it. Hence, the initial load-free bedrock $b^0$ cannot be subject to
a water load. At the grounding line, ice and water exert the same
pressure such that the bedrock extends continuously into its submerged
part.

Again we let the problem have the form
\begin{equation}
  M \Du{}{t} \bvec{x} = F(\bvec{x}, \lambda),  
\end{equation}
but now with $\bvec{x} = [\bvec{h}^T, \bvec{u}^T, \bvec{b}^T, x_g]^T$, $\bvec{h},\bvec{u},\bvec{b} \in \mathbb{R}^N$, $M \in \mathbb{R}^{(3N+1)\times(3N+1)}$ and $F : \mathbb{R}^{(3N+1)} \to \mathbb{R}^{(3N+1)}$, given by 
\begin{equation}
  M = \begin{bmatrix}~~I~ & 0 & 0 & M_{\text{mass}}(\bvec{h},x_g) \\ 0 & 0 & 0 & 0 \\ 0 & 0 & I & M_{\text{bed}}(\bvec{b},x_g)  \\ 0 & 0 & 0 & 0 \end{bmatrix}, 
\quad F = \begin{bmatrix} F_{\text{mass}}(\bvec{h},\bvec{u},\bvec{b},x_g) \\ F_{\text{mom}}(\bvec{h},\bvec{u},\bvec{b},x_g) \\ F_{\text{bed}}(\bvec{h},\bvec{b},x_g) \\ F_{\text{flot}}(\bvec{h},\bvec{b},x_g)\end{bmatrix}.
\end{equation}
The functions $M_{\text{mass}}$ and $M_{\text{bed}}$ give the
discretizations of the stretchings in the left hand side of the
continuity and bedrock equation, $F_{\text{bed}} \in \mathbb{R}^N$
gives the discretization of the elastic bedrock equation
\eqref{eq:bedrock2}. The other functions are the same as in Section
\ref{sec:implementation}, except with their dependence on $\bvec{b}$
made explicit.

Instead of having a constant accumulation $a$, as in
\cite{vdBerg_2006} we will give it a linear dependence on the surface
height $s(x) = h(x) - b(x)$ and a ceiling $a_\text{max}$:
\begin{equation}
a(s) = \min\left(a_\text{max},\theta(s-E)\right),
\end{equation}
where $\theta$ is an accumulation gradient and $E$ the equilibrium height. Below $E$ we have mass loss, i.e.~\emph{ablation}, the opposite of accumulation. The balance between accumulation and ablation is governed by air temperature, which we assume to decrease with increasing surface height. 

The implementation of the height-dependent accumulation is straightforward, mainly requiring a few extra dependencies on $h$ and $b$ in the Jacobian matrix. To obtain a smooth transition from the linear function $\theta(s-E)$ to the ceiling $a_\text{max}$, we use an approximation to the Heavyside function:

\begin{equation}
a(s) = a_\text{max} - \frac{1}{2} \left(1+\tanh\left(\frac{a_\text{max}-\theta(s-E)}{\epsilon}\right)\right)\left(a_\text{max}-\theta(s-E)\right), 
\end{equation}
with $\epsilon$ moderately small.

Recall the bedrock $b$ as given in \eqref{eq:initbedrock}. Now that we have implemented the isostatic adjustment, we need to define a smooth original bedrock $b^0$ that is not subject to a water load. Eventually, the bedrock will only be partially submerged, but adjusting $b$ only within the submerged sub-interval introduces unfavorable discontinuities. For that reason we will obtain $b^0$ using a global adjustment. 

Consider the stationary case of \eqref{eq:bedrock0}. Replacing the ice load $\rho_i h$ with a water load $\rho_w b$ gives the required adjustment:
\begin{equation}b^0 =  \frac{\rho_a-\rho_w}{\rho_a} b.\end{equation}

The added components open up a multitude of possible continuation parameters. Here we will restrict ourselves to a continuation in the equilibrium height $E$, to pursue oscillatory solutions due to the \emph{load accumulation feedback} \cite{Ghil_1994}: as ice thickness increases due to accumulation, the surface height may decrease as a result of isostatic adjustment (with a delay $\tau_a$), leading to a decrease in accumulation. This feedback suggests the possibility of oscillatory solutions within a certain parameter regime. 

\begin{figure}[htpb]
\includegraphics[clip=true,trim=5px 0px 0px 5px, width=.5\textwidth]{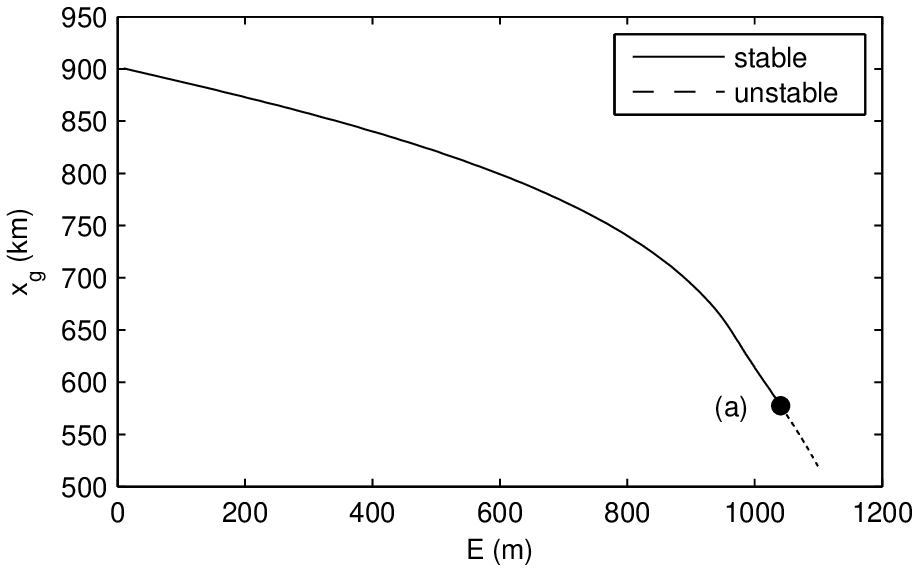}
\includegraphics[clip=true,trim=0px 0px 0px 0px, width=.5\textwidth]{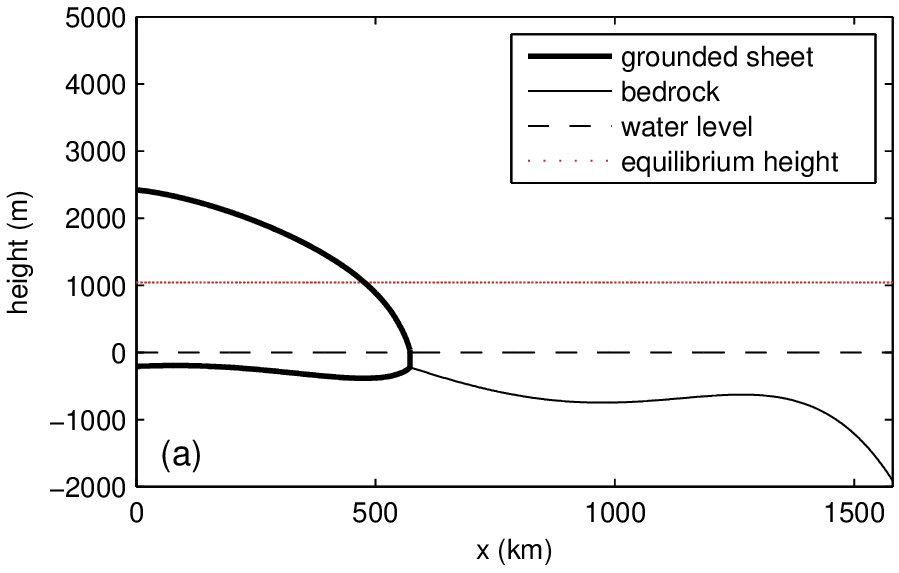}
\caption{Bifurcation diagram (left) and solution (right) of a continuation in equilibrium height $E$, starting at $E=0$. A Hopf bifurcation is detected at (a): $E=1.04 \times 10^3~\text{m}$ and the corresponding solution is plotted on the right. Instances of the oscillatory perturbation corresponding to the complex conjugate eigenpair are plotted in Figure \ref{fig:eigvechopf}. The number of grid-points is $N=800$, the other parameters are given in Table \ref{tab:sheethopf}. }\label{fig:sheethopf}
\end{figure}

A Hopf bifurcation occurs when a steady periodical solution emerges
from a fixed point. In the spectrum given by the eigenvalue analysis, 
a Hopf bifurcation corresponds to a
complex conjugate pair crossing the imaginary axis from the left to
the right half plane. The result of a continuation in $E$ is shown in
Figure \ref{fig:sheethopf}. An additional list of parameters is given
in Table \ref{tab:sheethopf}.

\begin{table}[htpb]\centering
  \begin{footnotesize}
  \begin{tabular}{|r|l|l|} 
    \hline \vspace{-2.5ex} &\\
 $\tau_a$           & $10000~\text{y}$ & asthenosphere relaxation timescale \\ 
 $\rho_a$           & $3300$~~$\text{kg}~\text{m}^{-3}$ & asthenosphere density \\
 $a_\text{max}$     & $0.1$~~$\text{m}~\text{y}^{-1}$   & maximum accumulation  \\     
 $\theta$           & $0.001~\text{y}^{-1} $ & accumulation gradient \\ 
 $A$                & $1.8969\times 10^{-26} ~\text{s}^{-1}~\text{Pa}^{-3}$ & Glen's flow law rheology parameter\\
\hline
  \end{tabular}\end{footnotesize}
  \caption{Parameter values for the experiment in Figure \ref{fig:sheethopf}. The accumulation ceiling and gradient are chosen similar to \cite{vdBerg_2006}. Parameters not mentioned here remain equal to the ones given in Table \ref{tab:rempars}.}
  \label{tab:sheethopf}
\end{table}
\begin{figure}[htbp]
\includegraphics[clip=true,trim=-3px 0px 0px 0px, width=.5\textwidth]{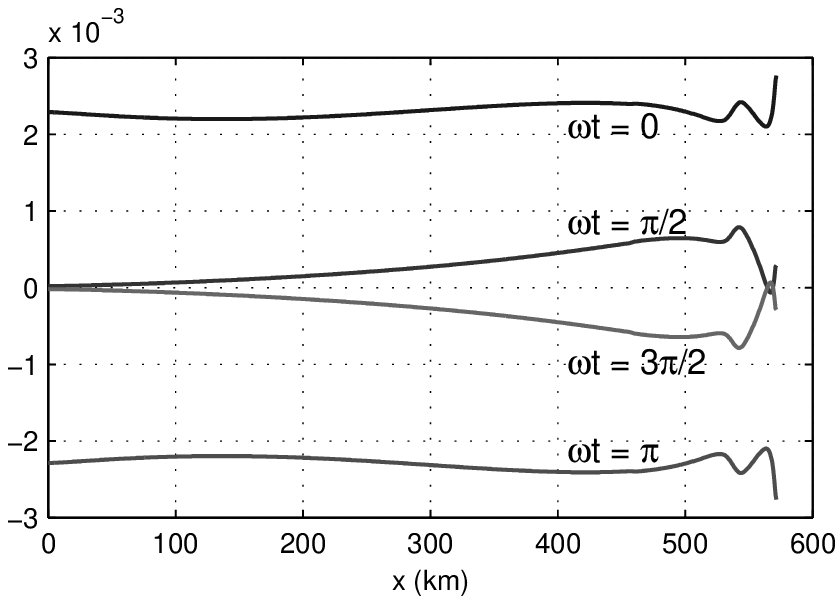}
\includegraphics[clip=true,trim=3px 0px 3px 3px, width=.5\textwidth]{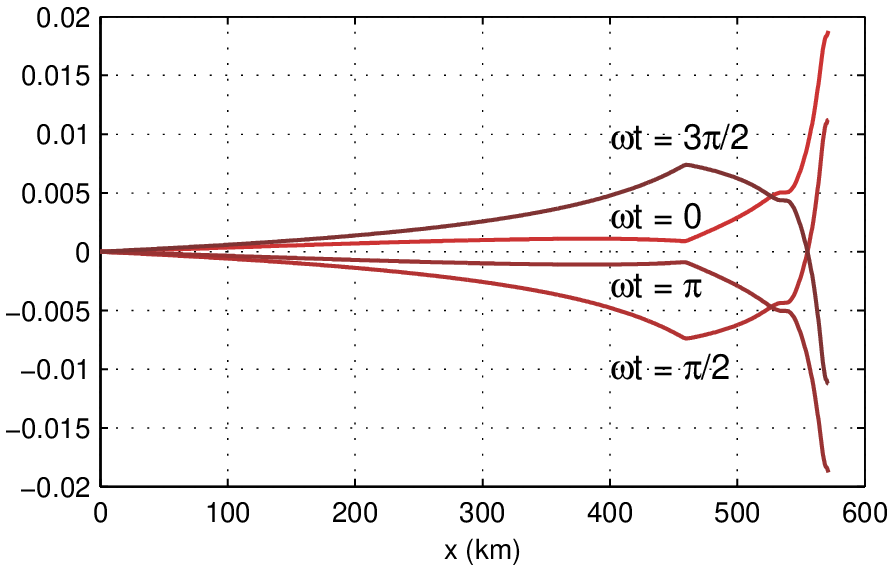}
\caption{Oscillating perturbation in ice thickness (left) and velocity (right), given by the eigenvector corresponding to the complex conjugate eigenpair at the Hopf bifurcation $E=1.04 \times 10^3~\text{m}$. Several instances of the perturbation are obtained using \eqref{eq:hopfpert}. The angular frequency is given by $\omega=\sigma_I$.} \label{fig:eigvechopf}
\end{figure}

A Hopf bifurcation is detected at $E=1.04 \times 10^3~\text{m}$, with complex conjugate eigenpair $\sigma = \sigma_R  \pm i \sigma_I  \approx 0 \pm i 0.006$. The corresponding complex eigenmode $\hat{x} = \hat{x}_R  \pm i \hat{x}_I$ describes the perturbation destabilizing the solution. The real part of the perturbation gives a disturbance profile \cite{Dijkstra_2005}: 
\begin{equation}
\text{Re}(\hat{\bvec{x}}\Exp{\sigma t}) = \Exp{\sigma_Rt}\left[\hat{x}_R\cos(\sigma_I t) - \hat{x}_I\sin(\sigma_I t) \right]. \label{eq:hopfpert}
\end{equation}
 At the bifurcation we have $\sigma_R = 0$, then $\hat{x}_R$ and $\hat{x}_I$ give two instances of the oscillatory perturbation: $\hat{x}_R$ at $\sigma_It=0$ and $\hat{x}_I$ at $\sigma_It = {3\pi}/{2}$, see Figure \ref{fig:eigvechopf}. The non-dimensional period is given by $\hat{T} = \frac{2\pi}{\sigma_I} \approx 1.047\times 10^3$. As the timescale in the experiment is taken $\tau_0 = 100~\text{y}$, the dimensional period is $T = 1.047\times 10^5~\text{y}$.

The obtained oscillating perturbation demonstrates the dynamics of the load accumulation feedback. For the ice thickness, the perturbation in the interior seems to be constantly ahead of the perturbation at the grounding line. Note that the region between the ice divide and the point at which the surface attains the equilibrium height ($x \approx 476~\text{km}$) is subject to a net accumulation. From the perturbation in thickness we see that the feedback is clearly driven by the accumulation. 
 
Beyond $x \approx 476~\text{km}$ there is a net ablation. In this region the perturbations in thickness and velocity show a drastic change in behavior. A steep increase in velocity occurs when the ice thickness decreases ($\omega t = \pi/2$), which corresponds to a large increase in flux to facilitate the mass loss. Similarly, when the sheet grows ($\omega t = 3\pi/2$), the flux needs to reduce to facilitate growth. 

\begin{figure}[htbp]\centering
\includegraphics[clip=true,trim=-3px 0px 0px 0px, width=.65\textwidth]{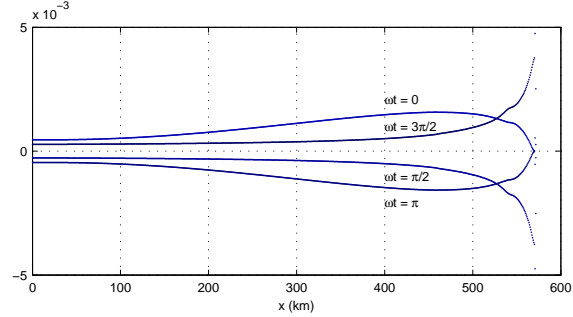}
\caption{Oscillating perturbation in the bedrock, given by the eigenvector corresponding to the complex conjugate eigenpair at the Hopf bifurcation.} \label{fig:eigvecbedrock}
\end{figure}

In the ablation region the perturbation in ice thickness shows a few peculiar oscillations. At the grounding line the sheet thickness seems to be slightly less than at the ice divide, perhaps to facilitate the appropriate flux. However, just before the grounding line and after $x \approx 476~\text{km}$ the effect of a reverse accumulation feedback seems to be visible. Due to the negative accumulation the bedrock is adjusted differently, see Figure \ref{fig:eigvecbedrock}. In the oscillatory bedrock perturbation there seems to be some irregular influence from the free boundary. Nevertheless, a change in adjustment around $x \approx 476~\text{km}$ is visible in at least two phases of the oscillation at $\omega t = 0$ and $\omega t = \pi$.

\section{Summary and Discussion} 

From the theories of ice-age cycles, it is clear that ice-sheet dynamics plays a central 
role in the explanation on how the variations in insolation lead to multi-millennial variability 
of the climate system, in particular on the 100 kyr time scale. A problem with the 
current theories is that there are many different conceptual models which can give a 
dominant 100 kyr variability, but it is difficult to falsify them based on the proxy 
data record \cite{Crucifix2016}. 

A step forward is  to determine spatial patterns of variability associated with the 
glacial cycles, similar to what has been done for other problems of climate 
variability such as El Ni\~no \cite{Dijkstra2002_ARFM} and the Atlantic Multidecadal 
Variability \cite{Frankcombe2010b}. This obviously requires a next level of 
models in the hierarchy (see Chapter 6 of \cite{Dijkstra2013B}), at least formulated
in terms of partial differential equations (spatially extended models), also 
(often) referred to as intermediate complexity models (ICMs). 

In this contribution, we applied techniques of numerical bifurcation theory to study the 
bifurcation behavior  of solutions of such an ICM  two-dimensional ice sheet 
model \cite{Schoof_2007}. The complication arises here from the grounding 
line dynamics at $x = x_g$, which is in principle a free boundary problem. Here, 
it is handled  by using a transformation $\transf = x/x_g$, where the  original 
domain $x \in [0,x_g]$  is mapped onto the fixed domain $\transf \in [0,1]$.  

In the version of the  model where the bottom topography is fixed, the results 
provide  insight into the  mechanism of  marine  ice sheet instability. There are 
robust intervals  in parameter space, where  multiple equilibria occur, 
corresponding to a large ice sheet and a small ice sheet. Hence, for these parameter 
values, transitions can occur where the ice-sheet decreases substantially in 
shape and length.  Tracing unstable equilibria and performing a stability  
analysis enables the  investigation of  the actual structure of the growing 
perturbation.   We have shown numerically  that a positive eigenvalue is
associated with the instability criterion \cite{Schoof_2012} for the
full problem, and that the advected thickness perturbation (the term
$\bar{u}\hat{h}$, where $\bar{u}$ is the steady state velocity and
$\hat{h}$ the ice thickness perturbation) dominates the instability
process. 

Transition behavior under the effects of noise in the accumulation 
parameter $a$ on the grounding line motion under stable conditions 
have been studied in \cite{Mulder2018}. The magnitude (for  typical 
noise levels) of these motions is in the order of 1000 m, which are similar amplitudes 
as observed for example in \cite{Hogg_2016}. It appears to be  more 
likely to jump from a large ice sheet state to a small ice sheet state 
than vice versa. Grounding line flux variability shows a related 
asymmetry, likely due to differences in local bedrock conditions 
and/or global ice sheet extent. 

When the load-accumulation feedback is included by extending the ice-sheet 
with a dynamical  bottom topography,  oscillatory instabilities occur through a 
Hopf bifurcation.  Here, the eigenvectors associated with the instability 
provide an interesting spatial pattern of variations of the ice sheet, with 
largest amplitudes near the grounding line. The time scale here is connected
to the relaxation time scale $\tau_a$ of the bottom topography reflecting 
the interaction between the ice sheet and bedrock  below. 

Of course, such an analysis does not solve the glacial - interglacial 
variability problem because the ICM used here does not capture essential 
processes. Currently, this land-ice model, a sea ice model and 
a carbon cycle model  are  coupled to the fully implicit ocean model 
THCM \cite{DeNiet2007, Thies2009} to provide a fully implicit climate 
model (still very idealized with respect to state-of-the-art global
climate models) with which bifurcation analysis can be performed. 
Using this model, we aim to show that there are different Hopf 
bifurcations involved in the glacial - interglacial problem and 
how the external M-forcing can interact with the internal variability 
generated by these instabilities. 

\begin{acknowledgement}
TEM and HAD acknowledge support by the Netherlands Earth System
Science Centre (NESSC), financially supported by the Ministry of
Education, Culture and Science (OCW), Grant no. 024.002.001. 
FWW acknowledge support from the Mathematics of Planet Earth 
research program, project number 657.000.007, which is financed 
by the Netherlands Organisation for Scientific Research (NWO).
\end{acknowledgement}

\bibliographystyle{spphys}
\bibliography{bif_IA_vfin}

\end{document}